\documentclass[12pt]{article}
\usepackage{graphicx}

\begin{document}

\begin{center}
{\Large{\bf Resonant Generation of Topological Modes in Trapped
Bose Gases} \\ [5mm]

V.I. Yukalov$^{1,2}$, K.-P. Marzlin$^{3,2}$, and E.P. Yukalova$^4$} \\ [5mm]

{\it
$^1$Bogolubov Laboratory of Theoretical Physics \\
Joint Institute for Nuclear Research, Dubna 141980, Russia\\ [3mm]
$^2$Fachbereich Physik, Universit\"at Konstanz, Fach M 674 \\
D-78457 Konstanz, Germany \\ [3mm]
$^3$Department of Physics and Astronomy\\
2500 University Drive NW, Calgary\\ 
Alberta T2N 1N4, Canada\\ [3mm]
$^4$Department of Computational Physics \\
Laboratory of Information Technologies \\
Joint Institute for Nuclear Research, Dubna 141980, Russia}

\end{center}

\vskip 2cm

\begin{abstract}

Trapped Bose atoms cooled down to temperatures below the Bose-Einstein
condensation temperature are considered. Stationary solutions to the
Gross-Pitaevskii equation (GPE) define the topological coherent modes,
representing nonground-state Bose-Einstein condensates. These modes can
be generated by means of alternating fields whose frequencies are in
resonance with the transition frequencies between two collective energy
levels corresponding to two different topological modes. The theory of
resonant generation of these modes is generalized in several aspects:
Multiple-mode formation is described; a shape-conservation criterion is
derived, imposing restrictions on the admissible spatial dependence
of resonant fields; evolution equations for the case of three coherent
modes are investigated; the complete stability analysis is accomplished;
the effects of harmonic generation and parametric conversion for the
topological coherent modes are predicted. All considerations are realized
both by employing approximate analytical methods as well as by numerically
solving the GPE. Numerical solutions confirm all conclusions following  
from analytical methods.

\end{abstract}

\vskip 2cm


\newpage

\section{Introduction}

Dilute Bose gases at low temperatures, when almost
all atoms are in Bose-Einstein condensate, are well described
by the GPE (see reviews
\cite{orgcit1}--\cite{orgcit5}). Since
the latter is a type of the nonlinear Schr\"odinger equation,
it should possess the whole spectrum of stationary solutions.
In the presence of a trapping potential, the related energy
spectrum is, in general, discrete. Stationary solutions to the
GPE are, by definition, the topological
coherent modes, whose ground state describes the standard
Bose-Einstein condensate, while the higher states correspond
to nonground-state Bose-Einstein condensates \cite{orgcit6}.

It is worth saying a few words recalling where the name {\it
topological coherent modes} comes from. Different stationary
solutions to the GPE, associated with
different energy levels, display principally different spatial
shapes, in particular, different number of zeros. Because of their
distinct spatial topology, the modes are termed {\it topological}.
These should not be confused with elementary collective excitations,
defined by linear Bogolubov - De Gennes equations. Such elementary
excitations describe small oscillations around a given stationary
solution and do not change the topology of the latter. Since the
GPE is nonlinear, its solutions can also be
named {\it nonlinear} modes, which would stress their principal
difference from elementary excitations satisfying the linear
Bogolubov - De Gennes equations. However, the elementary
excitations, produced by strong perturbations, are sometimes
also called nonlinear. Therefore the term topological,
characterizing dissimilar modes, seems to be more precise.
The topological modes are specified as {\it coherent} due to
the fact that the GPE can be interpreted
as an exact equation for coherent states \cite{orgcit7,orgcit8}.

The possibility of resonant generation of arbitrary topological
coherent modes was advanced in Ref. \cite{orgcit6}. A particular case
of vortex creation was suggested in \cite{orgcit9,orgcit10,orgcit11}.
The properties of such modes were also studied in \cite{orgcit12} --
\cite{orgcit22} and a dipole topological mode was excited in experiment
\cite{orgcit23}. Bose-Einstein condensates with topological coherent
modes exhibit a variety of interesting features which could find many
applications. Among these features, we could mention the following.
{\it Mode locking}: this is the effect under which the fractional
mode populations are locked to stay in the vicinity of their initial
values \cite{orgcit6,orgcit8,orgcit22}. Mathematically, this effect is  
analogous to self-trapping occurring for atoms in double-well potentials  
\cite{orgcit24,orgcit25,orgcit26}. {\it Critical dynamics}: an abrupt  
qualitative change of dynamics of fractional mode populations under an 
infinitesimally small variation of the pumping parameters
\cite{orgcit8,orgcit17,orgcit21,orgcit22}.
A mathematically similar effect in the case of
double-well potentials is the dynamic phase transition between
the Rabi and Josephson regimes
\cite{orgcit24,orgcit25,orgcit27}.
{\it Interference patterns}: specific interference fringes arising
because of differing spatial shapes of different topological modes
\cite{orgcit8,orgcit21}.
{\it Interference current}: fastly oscillating current, existing
even inside a single-well trap
\cite{orgcit8,orgcit21}. Such a type of current,
arising between two interpenetrating populations, not separated
by any barrier, is, on occasion, called the internal Josephson
effect
\cite{orgcit28,orgcit29}.
{\it Atomic squeezing}: narrowing of the dispersion
corresponding to the mode population difference, compared to the
dispersion related to dipole transitions \cite{orgcit8}. A similar
feature is illustrated by two-component condensates \cite{orgcit30}.
{\it Irreducible modes}: these are the topological modes that have no
linear counterparts, so that they cannot be considered as analytical
continuations, under increasing of nonlinearity, of the related
linear modes
\cite{orgcit18,orgcit19}. Such modes, in  addition to strong
nonlinearity, require the presence of multiwell potentials
\cite{orgcit31,orgcit32,orgcit33,orgcit34}.
An investigation of the mode spectrum in the case
of single-well traps shows that in this case nonlinear modes
are reducible and can be treated as analytical continuations
of linear counterparts
\cite{orgcit6,orgcit8,orgcit15,orgcit35,orgcit36}.

In the present paper, we generalize the theory of resonant
formation of topological coherent modes and study the features
that have not been considered in previous publications. The most
important novel points are as follows:

\begin{enumerate}

\item
The possibility of resonant generation of multiple topological
coherent modes is described. This can be achieved by subjecting
a trapped Bose-Einstein condensate to the action of several
alternating fields, whose frequencies are tuned to distinct
transition frequencies related to different modes 
(Sec.~\ref{sec:resGen}).

\item
A general criterion is derived, showing when nonlinear modes
cannot be generated even if the applied alternating field is in
perfect resonance with the corresponding transition frequency.
This condition relates the spatial dependence of the trapping
potential and that of the alternating field (Sec.~\ref{sec:nogo}).

\item
Simultaneous generation of two excited coherent modes is studied
in detail. The resulting dynamical system describes three coexisting
nonlinear modes (Sec.~\ref{sec:threeModes}).

\item
The phase portrait of the three-mode dynamical system is investigated.
All fixed points are found and their stability is analysed 
(Sec.~\ref{sec:stabil}).

\item
Harmonic generation of topological modes is shown to exist. This
effect is analogous to optical harmonic generation 
(Sec.~\ref{sec:harmgen} ).

\item
Parametric conversion for topological modes is another effect having
its optical counterpart. To realize this effect, it is necessary to
subject trapped atoms to the action of several alternating fields
(Sec.~\ref{sec:harmgen}).

\item
The principal feature of this article is that all consideration has
been done in two ways, by applying approximate analytical methods and
also by numerically solving the GPE. Numerical
solutions confirm all effects described analytically.

\end{enumerate}

\section{Resonant Generation of Multiple Modes}
\label{sec:resGen}
Dilute Bose-condensed gases at low temperature are characterized
\cite{orgcit1}--\cite{orgcit5}
by a coherent field $\varphi({\bf r},t)$, which is a wave function satisfying
the GPE
\begin{equation}
\label{org1}
i\hbar\; \frac{\partial}{\partial t} \; \varphi({\bf r},t) =
\left (\hat H[\varphi]+\hat V\right )\; \varphi({\bf r},t) \; ,
\end{equation}
where the nonlinear Hamiltonian
\begin{equation}
\label{org2}
\hat H[\varphi] = -\hbar^2 \; \frac{\nabla^2}{2m_0} + U({\bf r}) +
N\; A_s \; |\varphi({\bf r},t)|^2
\end{equation}
contains a trapping potential $U({\bf r})$ and the interaction intensity
\begin{equation}
\label{org3}
A_s \equiv 4\pi \hbar^2 \; \frac{a_s}{m_0} \; ,
\end{equation}
with $m_0$ being atomic mass; $a_s$, scattering length; and $N$, the
total number of atoms. The wave function is normalized to unity, so
that $||\varphi||=1$. The confining potential can be modulated by applying
an additional field $\hat V =V({\bf r},t)$.

The topological coherent modes are the solutions to the stationary
GPE
\begin{equation}
\label{org4}
\hat H[\varphi_n]\; \varphi_n({\bf r}) = E_n\; \varphi_n({\bf r}) \; ,
\end{equation}
where $n$ is a labelling multi-index. The transition frequencies for
two distinct modes $m\neq n$ are
\begin{equation}
\label{org5}
\omega_{mn} \equiv \frac{1}{\hbar}\; ( E_m - E_n) \; .
\end{equation}
The trap modulation is resonant if the frequency of an applied
alternating field is tuned close to one of the transition frequencies
(\ref{org5}). This resonant field can induce transitions between the considered
modes. Everywhere in what follows, talking about resonance, we keep in
mind the resonant transitions between distinct topological modes. This
should not be confused with parametric resonance, when one considers a
single mode, whose width can become divergent under special conditions
on the amplitude of a perturbing field \cite{orgcit37}.

Previously, the resonant excitation of topological modes has been
considered for the case of a sole resonant field coupling two chosen
modes
\cite{orgcit6,orgcit8}. Now, we turn to the most general case, when there are
several alternating fields, so that the modulating potential is a sum
\begin{equation}
\label{org6}
V({\bf r},t) = \sum_j \left [ V_j({\bf r})\; \cos(\omega_j t) +
V_j'({\bf r})\; \sin(\omega_j t) \right ]
\end{equation}
of fields with different frequencies $\omega_j$. Generally, the amplitudes
$V_j({\bf r})$ and $V_j'({\bf r})$ can also be slow functions of time,
whose temporal variation is slow compared to that of $\cos(\omega_jt)$,
such that
\begin{equation}
\label{org7}
\frac{1}{\hbar\omega_j} \; \left | \frac{\partial V_j}{\partial t}
\right | \ll 1 \; , \qquad
\frac{1}{\hbar\omega_j} \; \left | \frac{\partial V_j'}{\partial t}
\right | \ll 1 \; .
\end{equation}
If these amplitudes are such slow functions of time, then, varying them  
adiabatically, one could induce the Landau-Zener tunneling
\cite{orgcit31,orgcit33}. But here we shall consider
only the transitions caused by the fastly oscillating $\cos(\omega_j t)$.
Each frequency $\omega_j$ is assumed to be close to one of the transition
frequencies (\ref{org5}), so that the {\it resonance conditions}
\begin{equation}
\label{org8}
\left | \frac{\Delta_{mn}}{\omega_{mn}}\right | \ll 1 \; , \qquad
\Delta_{mn} \equiv \omega_j - \omega_{mn} \; ,
\end{equation}
are valid.

The modulating potential (\ref{org6}) can be presented as
\begin{equation}
\label{org9}
V({\bf r},t) = \frac{1}{2} \; \sum_j \left [ B_j({\bf r})\; e^{i\omega_j t} +
B_j^*({\bf r})\; e^{-i\omega_j t} \right ] \; ,
\end{equation}
with
\begin{equation}
\label{org10}
B_j({\bf r}) \equiv V_j({\bf r}) - i V_j'({\bf r}) \; .
\end{equation}
The modulation can be realized by varying the trapping magnetic fields,
by invoking laser spoons, or resorting to other means
\cite{orgcit6,orgcit8,orgcit9,orgcit10,orgcit11}.

There exist two characteristic quantities, called transition amplitudes;
one is the matrix element
\begin{equation}
\label{org11}
\alpha_{mn} \equiv N\; \frac{A_s}{\hbar} \; \left (
|\varphi_m|^2,\; 2|\varphi_n|^2-|\varphi_m|^2 \right ) \; ,
\end{equation}
due to the interatomic interactions (\ref{org3}), and another one is the matrix
element
\begin{equation}
\label{org12}
\beta_{mn} \equiv \frac{1}{\hbar} \; \left (
\varphi_m,\hat B_j\varphi_n\right ) \; ,
\end{equation}
related to the amplitude $\hat B_j = B_j({\bf r})$ of the modulating
potential (\ref{org9}). Here, $(. , .)$ denotes the usual scalar
product of Schr\"odinger theory.
To avoid intensive power broadening, these amplitudes
(\ref{org11}) and (\ref{org12}) have to satisfy the inequalities
\begin{equation}
\label{org13}
\left | \frac{\alpha_{mn}}{\omega_{mn}}\right | \ll 1 \; , \qquad
\left | \frac{\beta_{mn}}{\omega_{mn}}\right | \ll 1 \; ,
\end{equation}
where $m\neq n$ and whose meaning was explained in detail in \cite{orgcit8}.
Conditions (\ref{org13}) allow for an effective generation of nonlinear modes
by resonant fields. The first of these restrictions, briefly speaking,
can be reduced to the limitation on the number of atoms that can be
transferred to an excited coherent mode \cite{orgcit8}. This limiting number
of atoms is close to the critical number of atoms with attractive
interactions, for which the Bose-Einstein condensate preserves its
stability
\cite{orgcit6,orgcit8,orgcit38,orgcit39}.
Therefore the resonant generation of nonlinear
modes is feasible for atoms with positive as well as negative
scattering lengths.

The topological coherent modes, being the solutions to the
nonlinear eigenproblem (\ref{org4}), do not compulsory form a complete
orthonormal basis. However, the modes $\varphi_n({\bf r})$ can always be
normalized, so that $||\varphi_n||=1$. And the set $\{\varphi_n({\bf r})\}$
of all linearly independent functions forms a total basis \cite{orgcit8},
permitting one to look for the solution of the Gross-Pitaevskii
equation (\ref{org1}) in terms of the presentation
\begin{equation}
\label{org14}
\varphi({\bf r},t) = \sum_n c_n(t)\; \varphi_n({\bf r}) \; \exp\left ( -\;
\frac{i}{\hbar}\; E_n\; t\right ) \; ,
\end{equation}
where $c_n(t)$ are unknown functions of time. Note that for some
nonlinear eigenproblems, it has been rigorously proved
\cite{orgcit40,orgcit41,orgcit42}
that the set of the corresponding eigenfunctions forms a complete
basis. In our case, it is sufficient to require that the functions
$c_n(t)$ are slowly varying, such that
\begin{equation}
\label{org15}
\frac{1}{\omega_{mn}}\; \left | \frac{dc_n}{dt} \right | \ll 1 \; ,
\end{equation}
and conditions (\ref{org13}) hold true. Then the expansion (\ref{org14})
is uniquely defined by means of the averaging technique \cite{orgcit43}.

It is worth emphasizing that the presentation (14) ideally suits for
analysing resonant transitions between the coherent topological modes.
And it is solely these modes that are the subject of the present paper.
We shall not consider here other possible excitations that could be  
produced by nonresonant driving fields and studied by means of the known  
rescaling procedure. Nontopological nonresonant breathing-type 
oscillations have been considered by many authors in the early days of
the BEC research (see reviews \cite{orgcit1}--\cite{orgcit5}). Therefore
there is no reason of extending the paper by repeating similar results.
This is why here we limit ourselves by treating only the resonant 
generation of topological modes, which have not been studied earlier.

Aiming at exciting particular modes, one should keep in mind, in
addition to the resonance conditions (\ref{org8}), the symmetry properties
of the corresponding wave functions, for which the modulating
field has to be such that the transition amplitudes (\ref{org12}) be nonzero.
However, even if all above mentioned conditions hold true, there
exists a rather general situation when the generation of modes
is impossible.

\section{No-Go Theorem for Mode Generation}
\label{sec:nogo}
It may happen that the applied modulating field is not able to generate
higher nonlinear modes, but can only lead to the oscillation of the
initial wave function, without changing its shape. More precisely, let
us start with a wave function $\varphi({\bf r},0)$. After a modulating field
$V({\bf r},t)$ begins acting on atoms, the function $\varphi({\bf r},0)$
is transferred to a function $\varphi({\bf r},t)$. When the {\it
shape-conservation condition}
\begin{equation}
\label{org16}
|\varphi({\bf r},t)| = |\varphi({\bf r}-{\bf a},0)| \; ,
\end{equation}
holds true for time dependent 
${\bf a}={\bf a}(t)$, 
then the shape of the atomic cloud
does not change in time, but the cloud oscillates as a whole, with
its center of mass moving according to the dependence ${\bf a}(t)$.
Hence, if we start with a mode $\varphi({\bf r},0)$ it can never be
transferred to another mode.

By definition, the initial function $\varphi({\bf r},0)=\varphi_0({\bf r})$
presents a nonlinear mode if it satisfies the stationary Gross-Pitaevskii
equation
\begin{equation}
\label{org17}
\hat H[\varphi_0({\bf r})]\;\varphi_0({\bf r}) = E_0\;\varphi_0({\bf r}) \; ,
\end{equation}
with the nonlinear Hamiltonian (\ref{org2}). 
We assume that this initial mode is real-valued, i.e., 
\begin{equation}
\label{org18}
\varphi({\bf r},0) =\varphi_0({\bf r}) =\varphi_0^*({\bf r}) \; .
\end{equation}
An example for this would be the ground-state
of a Bose-Einstein condensate.
In addition, we are focusing on {\it trapped} atoms, 
which implies that the confining potential $U({\bf r})$
increases towards infinity for $r\equiv|{\bf r}| \rightarrow \infty$. 
Therefore, the {\it trapping condition}
\begin{equation}
\label{org19}
\lim_{{\bf r}\rightarrow\infty} \; \varphi({\bf r},t) = 0
\end{equation}
is valid for all $t\geq 0$.

\vskip 2mm

{\it Theorem}. Suppose that atoms in a trapping potential $U({\bf r})$,
being initially in a real mode $\varphi_0({\bf r})$, are subject to the
action of a modulating field $V({\bf r},t)$, so that conditions (\ref{org17}) to
(\ref{org19}) are valid. Then the solution of the temporal Gross-Pitaevskii
equation (\ref{org1}) preserves the shape of the initial mode, satisfying
condition (\ref{org16}), if and only if the trapping potential is harmonic,
\begin{equation}
\label{org20}
U({\bf r}) = A_0 +{\bf A}_1\cdot{\bf r} +\sum_{\alpha\beta} A_{\alpha\beta}\;
r^\alpha\; r^\beta \; ,
\end{equation}
where $\alpha$ and $\beta$ are the Cartesian indices,
while the modulating field is linear with respect to the spatial
variables,
\begin{equation}
\label{org21}
V({\bf r},t) = B_0(t) +{\bf B}_1(t)\cdot{\bf r} \; ,
\end{equation}
$B_0(t)$ and ${\bf B}_1(t)$ being arbitrary functions of time. And the
center-of-mass motion ${\bf a}={\bf a}(t)$ is described by the equation
\begin{equation}
\label{org22}
m_0 \; \frac{d^2a^\alpha}{dt^2} + \sum_\beta \left ( A_{\alpha\beta} +
A_{\beta\alpha}\right )\; a^\beta + B_1^\alpha(t) = 0 \; .
\end{equation}

\vskip 2mm

{\it Proof}. Assume that the shape-conservation condition (\ref{org16})
holds true, and let us show that then the trapping and modulating
potentials are necessarily such as in Eqs. (\ref{org20}) and (\ref{org21}). For this
purpose, it is useful to invoke the presentation
\begin{equation}
\label{org23}
\varphi({\bf r},t) = |\varphi({\bf r},t)|\; \exp\{ i S({\bf r},t)\} \; ,
\end{equation}
where $S({\bf r},t)$ is a real action defining the velocity
\begin{equation}
\label{org24}
{\bf v}({\bf r},t) \equiv \frac{\hbar}{m_0}\; \vec\nabla S({\bf r},t) \; .
\end{equation}
Substituting the presentation (\ref{org23}) into the GPE
(\ref{org1}) yields
\cite{orgcit1}--\cite{orgcit5} the continuity equation

\begin{equation}
\label{org25}
\frac{\partial}{\partial t} \; |\varphi|^2 +\vec\nabla(|\varphi|^2{\bf v}) = 0
\end{equation}
and the velocity equation
\begin{equation}
\label{org26}
m_0\; \frac{\partial{\bf v}}{\partial t} = -\vec\nabla U_{eff} \; ,
\end{equation}
in which the effective potential
\begin{equation}
\label{org27}
U_{eff} = U({\bf r}) + V({\bf r},t) + N\; A_s\; |\varphi|^2 - \;
\frac{\hbar^2\vec\nabla^2|\varphi|}{2m_0|\varphi|} + \frac{m_0v^2}{2}\; .
\end{equation}
Because of the shape-conservation condition (\ref{org16}), one has
$$
\frac{\partial|\varphi|}{\partial t} = -\vec\nabla|\varphi|\cdot
\frac{d{\bf a}}{dt} \; .
$$
Using this in the continuity equation (\ref{org25}) results in
$$
\vec\nabla\left [ |\varphi|^2 \left ({\bf v}-\;
\frac{d{\bf a}}{dt}\right ) \right ] = 0 \; ,
$$
from where
$$
{\bf v} = \frac{d{\bf a}}{dt} + \frac{{\bf c}}{|\varphi|^2} \; ,
\qquad {\bf c}={\bf c}(t) \; .
$$
If here the function ${\bf c}(t)$ is not zero, then from the trapping
condition (\ref{org19}) it follows that ${\bf v}\rightarrow\infty$ as $r\rightarrow\infty$ hence,
the action $S\rightarrow\infty$ as $r\rightarrow\infty$. But then the limit $r\rightarrow\infty$
of the function (\ref{org23}) is not defined. Therefore, ${\bf c}(t)=0$. Thus, the
velocity (\ref{org24}) becomes
\begin{equation}
\label{org28}
{\bf v} = \frac{d{\bf a}}{dt} ={\bf v}(t) \; ,
\end{equation}
which is a function of time only. But then the velocity equation (\ref{org26}) tells
us that $\vec\nabla U_{eff}$ is also a function of time only, hence the
effective potential is linear in ${\bf r}$, having the form
\begin{equation}
\label{org29}
U_{eff} = D_0(t) +{\bf D}_1(t)\cdot{\bf r} \; .
\end{equation}
Taking into account that the initial mode is real-valued, as is conditioned
by Eq. (\ref{org18}), and employing the shape-conservation condition (\ref{org16}), one gets
\begin{equation}
\label{org30}
|\varphi({\bf r},t)| =\varphi_0({\bf r}-{\bf a}) \; .
\end{equation}
Then the effective potential (\ref{org27}) can be written as
\begin{equation}
\label{org31}
U_{eff} = U({\bf r}) - U({\bf r} -{\bf a}) + V({\bf r},t) + E_0 +
\frac{m_0v^2}{2} \; .
\end{equation}
Since, according to Eq. (\ref{org29}), the effective potential is linear in ${\bf r}$,
$E_0$ is a constant, and $v=v(t)$ is a function of time, then the sum of
the first three terms in Eq. (\ref{org31}) should also be linear in ${\bf r}$, such
that
\begin{equation}
\label{org32}
U({\bf r}) - U({\bf r}-{\bf a}) + V({\bf r},t) = U_0(t) + {\bf U}_1(t)\cdot{\bf r} \; .
\end{equation}
Thence the effective potential (\ref{org31}) takes the form
\begin{equation}
\label{org33}
U_{eff} = E_0 + \frac{m_0v^2}{2} + U_0(t) + {\bf U}_1(t)\cdot{\bf r} \; .
\end{equation}
Comparing Eqs. (\ref{org29}) and (\ref{org33}), we have
\begin{equation}
\label{org34}
D_0(t) = E_0 + \frac{m_0v^2}{2} + U_0(t) \; , \qquad
{\bf D}_1(t) ={\bf U}_1(t) \; .
\end{equation}
Equality (\ref{org32}) can be satisfied only if the trapping and modulating
potentials are given by Eqs. (\ref{org20}) and (\ref{org21}). Substituting
the latter in Eq. (\ref{org32}), we find
$$
U_0(t) = B_0(t) +{\bf a}\cdot{\bf A} - \sum_{\alpha\beta}
A_{\alpha\beta}\; a^\alpha\; a^\beta \; ,
$$
\begin{equation}
\label{org35}
U_1^\alpha(t) = B_1^\alpha(t) + \sum_\beta \left ( A_{\alpha\beta} +
A_{\beta\alpha} \right )\; a^\beta \; .
\end{equation}
Combining Eqs. (\ref{org26}), (\ref{org28}), (\ref{org33}), and
(\ref{org35}), we obtain Eq. (\ref{org22}) for the center-of-mass motion.

Now let us show that Eqs. (\ref{org20}) and (\ref{org21}) are sufficient
for the validity of the shape-conservation condition (\ref{org16}). Equations
(\ref{org25}) and (\ref{org26}), with the effective potential (\ref{org27})
and with the initial conditions
$$
|\varphi({\bf r},0)| =\varphi_0({\bf r}) \; , \qquad
{\bf v}({\bf r},0) = 0 \; ,
$$
are equivalent to the GPE (\ref{org1}) with the initial
condition (\ref{org18}). The latter equation is a nonlinear Schr\"odinger
equation, which, being complimented by the boundary condition (\ref{org19}),
possesses a unique solution \cite{orgcit44}. Hence, Eqs. (\ref{org25}) and
(\ref{org26}), with the same boundary condition (\ref{org19}), enjoy a unique
solution. These equations, under conditions (\ref{org20}) and (\ref{org21}),
do have a solution satisfying the shape-conservation condition (\ref{org16}),
which, according to the aforesaid, is a unique solution. This concludes the
proof.

\vskip 2mm

One should not confuse the shape-conservation criterion derived here with
the known result of the decoupled center-of-mass motion in a harmonic
potential. The criterion of shape conservation shows when the wave function
retains its shape under the action of an external alternating field and when
the shape is not preserved. The center-of-mass motion is a trivial byproduct
of our theorem. The principal point is the shape conservation. The derived
criterion shows that the trapping potential may be harmonic, but, if the
driving field is nonlinear, then, irrespectively of the center-of-mass motion,
the shape of the wave function will not be conserved.

This theorem teaches us that if the trapping potential is harmonic, which
is a standard situation, then  the modulating field, being linear in 
${\bf r}$, is not able to generate topological modes, even if its 
alternating temporal parts oscillate in an exact resonance with the 
corresponding transition frequencies. It is in full agreement with a 
nonlinear Ehrenfest-Theorem for the mean position and variance of the 
collective wavefunction \cite{orgcit11}. To generate such modes, at least 
one of the conditions (\ref{org20}) or (\ref{org21}) has to be broken.
But if a resonant alternating field is linear and the trapping potential 
is harmonic, then the real initial mode just moves in space, without 
changing its shape. We have checked this conclusion  by numerically 
solving the GPE under conditions (\ref{org20}) 
and  (\ref{org21}) and found a perfect agreement with the theorem.

\section{Coupling of Three Nonlinear Modes}
\label{sec:threeModes}
Several modes can be generated by applying a modulating field (\ref{org6})
containing several corresponding resonant terms. In general, two cases
are admissible: when the excited modes are not coupled to each other and
when they are coupled. In the former case, the overall dynamics consists
of the motion of several pairs of modes, each pair being separated in its
motion from other modes. The motion of such separate pairs of resonant modes
has been studied in detail earlier
\cite{orgcit6,orgcit8,orgcit9,orgcit10,orgcit11,orgcit14,orgcit17,orgcit21,orgcit22}.
Therefore we now concentrate on the case of several coupled modes.

We consider the case of three coupled modes, whose transition frequencies
(\ref{org5}) are enumerated as $\omega_{21}$, $\omega_{31}$, and $\omega_{32}$,
keeping in mind that the related energy levels are such that $E_1<E_2<E_3$.
To couple the modes, it is sufficient to have two modulating fields of the
possible three that would be in resonance with the corresponding transition
frequencies. In general, there can be three detunings $\Delta_{21}$,
$\Delta_{31}$, and $\Delta_{32}$ satisfying the resonance conditions
(\ref{org8}). For the transition amplitudes (\ref{org12}), we have
$\beta_{12}$, $\beta_{13}$, and $\beta_{23}$. There are six amplitudes
(\ref{org11}) for $\alpha_{ij}$ with $i\neq j$.

Realizing the coupling of three modes by two driving fields, we can
create three types of resonant systems, which may be called, by analogy
with the similar situations for resonant atoms, as cascade, $V$-type,
and $\Lambda$-type systems \cite{orgcit45}. In the cascade-type generation,
the modes with the transition frequencies $\omega_{21}$ and $\omega_{32}$
are coupled. In the $V$-type case, the transition frequencies are
$\omega_{21}$ and $\omega_{31}$. And for the $\Lambda$-type system, the
transition frequencies are $\omega_{31}$ and $\omega_{32}$.

Substituting into the GPE (\ref{org1}) the presentation
(\ref{org14}) and the corresponding driving fields (\ref{org6}), we employ
the averaging technique \cite{orgcit43}. This procedure is absolutely
the same as has been thoroughly described in
\cite{orgcit6,orgcit8}, so we do not repeat it here. The result
is the system of equations for the coefficient functions
$$
i\; \frac{dc_1}{dt} =\left ( \alpha_{12}|c_2|^2+ \alpha_{13}|c_3|^2
\right )\; c_1 + F_1 \; ,
$$
$$
i\; \frac{dc_2}{dt} =\left ( \alpha_{21}|c_1|^2+ \alpha_{23}|c_3|^2
\right )\; c_2 + F_2 \; ,
$$
\begin{equation}
\label{org36}
i\; \frac{dc_3}{dt} =\left ( \alpha_{31}|c_1|^2+ \alpha_{32}|c_2|^2
\right )\; c_3 + F_3\; ,
\end{equation}
where the terms $F_i$ depend on the type of the generation method.
For the cascade generation, we have
$$
F_1 = \frac{1}{2}\; \beta_{12}\; c_2\; e^{i\Delta_{21}t} \; ,
$$
$$
F_2 = \frac{1}{2}\; \beta_{12}^*\; c_1\; e^{-i\Delta_{21}t} +
\frac{1}{2}\; \beta_{23}\; c_3\; e^{i\Delta_{32}t} \; ,
$$
\begin{equation}
\label{org37}
F_3 = \frac{1}{2}\; \beta_{23}^*\; c_2\; e^{-i\Delta_{32}t} \; .
\end{equation}
If we set here $\alpha_{13}$, $\alpha_{31}$, $\alpha_{23}$, $\alpha_{32}$, and
$\beta_{23}$ to zero, we return to the studied earlier two-mode case
\cite{orgcit6,orgcit8}. In the case of the $V$-type coupling,
$$
F_1 = \frac{1}{2}\; \beta_{12}\; c_2\; e^{i\Delta_{21}t} +
\frac{1}{2}\; \beta_{13}\; c_3\; e^{i\Delta_{31}t} \; ,
$$
\begin{equation}
\label{org38}
F_2 = \frac{1}{2}\; \beta_{12}^*\; c_1\; e^{-i\Delta_{21}t} \; , \qquad
F_3 = \frac{1}{2}\; \beta_{13}^*\; c_1\; e^{-i\Delta_{31}t} \; .
\end{equation}
And for the $\Lambda$-type generation,
$$
F_1 = \frac{1}{2}\; \beta_{13}\; c_3\; e^{i\Delta_{31}t} \; , \qquad
F_2 = \frac{1}{2}\; \beta_{23}\; c_3\; e^{i\Delta_{32}t} \; ,
$$
\begin{equation}
\label{org39}
F_3 = \frac{1}{2}\; \beta_{13}^*\; c_1\; e^{-i\Delta_{31}t} +
\frac{1}{2}\; \beta_{23}^*\; c_2\; e^{-i\Delta_{32}t} \; .
\end{equation}
In addition, from the normalization of the function (\ref{org14}), we have
\begin{equation}
\label{org40}
|c_1|^2 + |c_2|^2 + |c_3|^2 =  1\; .
\end{equation}
Each $c_i=c_i(t)$ defines the dynamics of the corresponding
fractional mode population $|c_i|^2$.

It is important to stress that Eqs. (\ref{org36}) are obtained by employing
the standard averaging method \cite{orgcit43}), taking into account the
existence of two time scales, slow and fast, related to the inequalities
(\ref{org13}) and (\ref{org15}). In this averaging procedure, one
substitutes expansion (\ref{org14}) into the GPE
(\ref{org1}) and averages over time fastly oscillating functions. As a
result of this, the so-called time-reversed or counter-rotating terms
vanish, so that in the right-hand side of Eqs. (\ref{org36}) there are no
terms like $c_1^*c_2^2$ or $c_2^*c_3^2$. It is well known that, even if  
such terms would be added, they do not produce a significant change in the 
solutions. Omitting these terms is essentially equivalent to the widely 
known rotating-wave approximation \cite{orgcit45}. The derivation of 
equations for $c_n(t)$, with all related mathematical details has been  
thoroughly described in Refs. \cite{orgcit6,orgcit8,orgcit22}.

Though Eqs. (\ref{org36}) look differently for different types of mode  
generation related to distinct terms $F_i$, the mathematical structure of 
these equations is, actually, the same. We may notice the following 
symmetry properties. The $V$-type equations can be obtained from the  
cascade-type ones by interchanging the indices 1 and 2 and by replacing 
$\beta_{21}$ by $\beta_{12}^*$. Similarly, the $\Lambda$-type equations  
can be derived from the cascade-type equations by interchanging the  
indices 2 and 3, with the replacement $\beta_{32}\rightarrow\beta_{23}^*$.  
The relation $\Delta_{ij}=-\Delta_{ji}$ has to be taken into account.  
Because of this symmetry, it is sufficient to consider just one type of  
Eqs. (\ref{org36}), for instance, that corresponding to the cascade 
generation.

The functions $c_i(t)$ are complex-valued. Hence, Eqs. (\ref{org36}) present
a system of six differential equations. However, it is possible to show that
they define a four-dimensional dynamical system. To prove this, we involve the
notation
\begin{equation}
\label{org41}
c_j =|c_j|\; \exp(i\pi_j) \; ,
\end{equation}
where $\pi_j=\pi_j(t)$ is a real-valued phase. Also, we write
\begin{equation}
\label{org42}
\beta_{ij} = b_{ij}\; \exp(i\gamma_{ij})\; , \qquad
b_{ij}\equiv |\beta_{ij}| \; .
\end{equation}
Introduce the population differences
\begin{equation}
\label{org43}
s\equiv |c_2|^2 - |c_1|^2 \; , \qquad p\equiv |c_3|^2 - |c_2|^2
\end{equation}
and the relative phases
\begin{equation}
\label{org44}
x\equiv \pi_2 -\pi_1 +\gamma_{12} + \Delta_{21} t \; , \qquad
y\equiv \pi_3 -\pi_2 +\gamma_{23} + \Delta_{32} t \; .
\end{equation}
The fractional mode populations can be expressed through the
variables (\ref{org43}) as
$$
|c_1|^2 = \frac{1}{3}\; ( 1 - 2s - p) \; , \qquad
|c_2|^2 = \frac{1}{3}\; ( 1 + s - p) \; \qquad
|c_3|^2 = \frac{1}{3}\; ( 1 + s + 2p) .
$$
Then Eqs. (\ref{org36}) for the cascade generation can be reduced to the
system of four equations
$$
\frac{ds}{dt} = \frac{1}{3}\; \sqrt{1+s-p}
\left ( b_{23}\; \sqrt{1+s+2p}\; \sin y - 2b_{12}\;
\sqrt{1-2s-p}\; \sin x\right ) \; ,
$$
$$
\frac{dp}{dt} = \frac{1}{3}\; \sqrt{1+s-p}
\left ( b_{12}\; \sqrt{1-2s-p}\; \sin x - 2b_{23}\;
\sqrt{1+s+2p}\; \sin y\right ) \; ,
$$
$$
\frac{dx}{dt} = \alpha_1 s + \delta_1 p +
\frac{3b_{12}\;s\cos x}{2\sqrt{(1+s-p)(1-2s-p)}} - \frac{1}{2}\; b_{23}\;
\sqrt{\frac{1+s+2p}{1+s-p}}\; \cos y + \delta_2 \; ,
$$
\begin{equation}
\label{org45}
\frac{dy}{dt} = \alpha_2 p + \delta_3 s +
\frac{3b_{23}\; p\cos y}{2\sqrt{(1+s-p)(1+s+2p)}} + \frac{1}{2}\; b_{12}\;
\sqrt{\frac{1-2s-p}{1+s-p}}\; \cos x + \delta_4 \; ,
\end{equation}
in which
$$
\alpha_1 \equiv \frac{1}{3}\;
( \alpha_{12} + \alpha_{13} + 2\alpha_{21} -\alpha_{23} ) \; , \qquad
\alpha_2 \equiv \frac{1}{3}\;
( \alpha_{32} + \alpha_{31} + 2\alpha_{23} -\alpha_{21} ) \; ,
$$
$$
\delta_1 \equiv \frac{1}{3}\;
( \alpha_{21} - \alpha_{12} + 2\alpha_{13} - 2\alpha_{23} ) \; , \qquad
\delta_2 \equiv \Delta_{21} + \frac{1}{3}\;
( \alpha_{12} - \alpha_{21} + \alpha_{13} -\alpha_{23} ) \; ,
$$
$$
\delta_3 \equiv \frac{1}{3}\;
( \alpha_{23} - \alpha_{32} + 2\alpha_{31} - 2\alpha_{21} ) \; , \qquad
\delta_4 \equiv \Delta_{32} + \frac{1}{3}\;
( \alpha_{23} - \alpha_{32} + \alpha_{21} -\alpha_{31} ) \; .
$$
Thus, the set of six equations (\ref{org36}) is really equivalent to
a four-dimensional dynamical system (\ref{org45}).

\begin{figure}
\centerline{
\includegraphics[width=9cm]{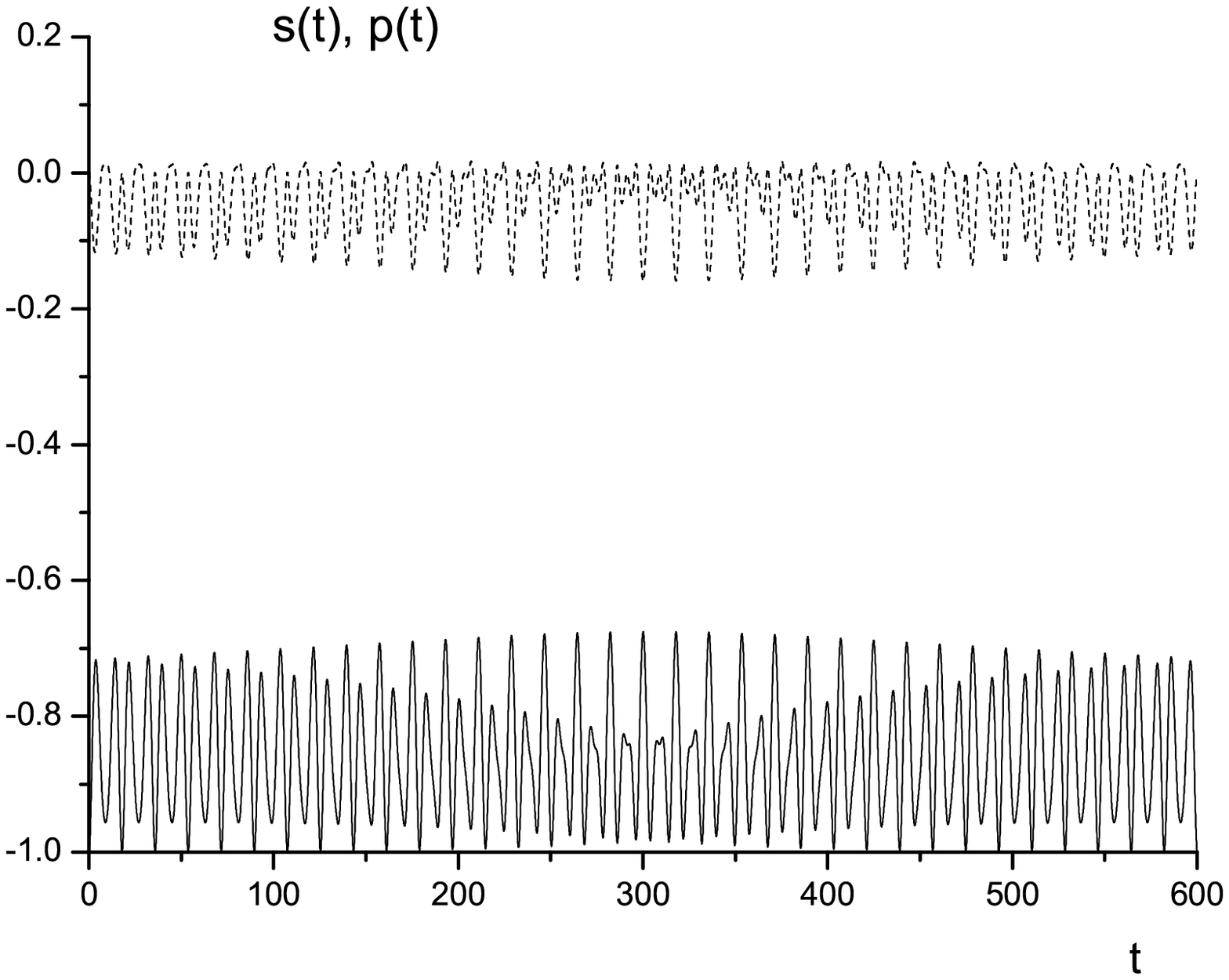}
}
\caption{\label{timeEvolSlavaFig1} Mode locked regime for the case of three
coupled nonlinear modes. Parameters are $\alpha_{ij}=\alpha$,
$b_{ij}=0.35\alpha$, and $\delta_i=0$. Initial conditions are
$s_0=-1$, $p_0=0$, $x_0=y_0=0$. Time is measured in units
of $\alpha^{-1}$. Shown are the population differences $s(t)$
(solid line) and $p(t)$ (dashed line).}
\end{figure}

\begin{figure}
\centerline{
a) \includegraphics[width=7cm]{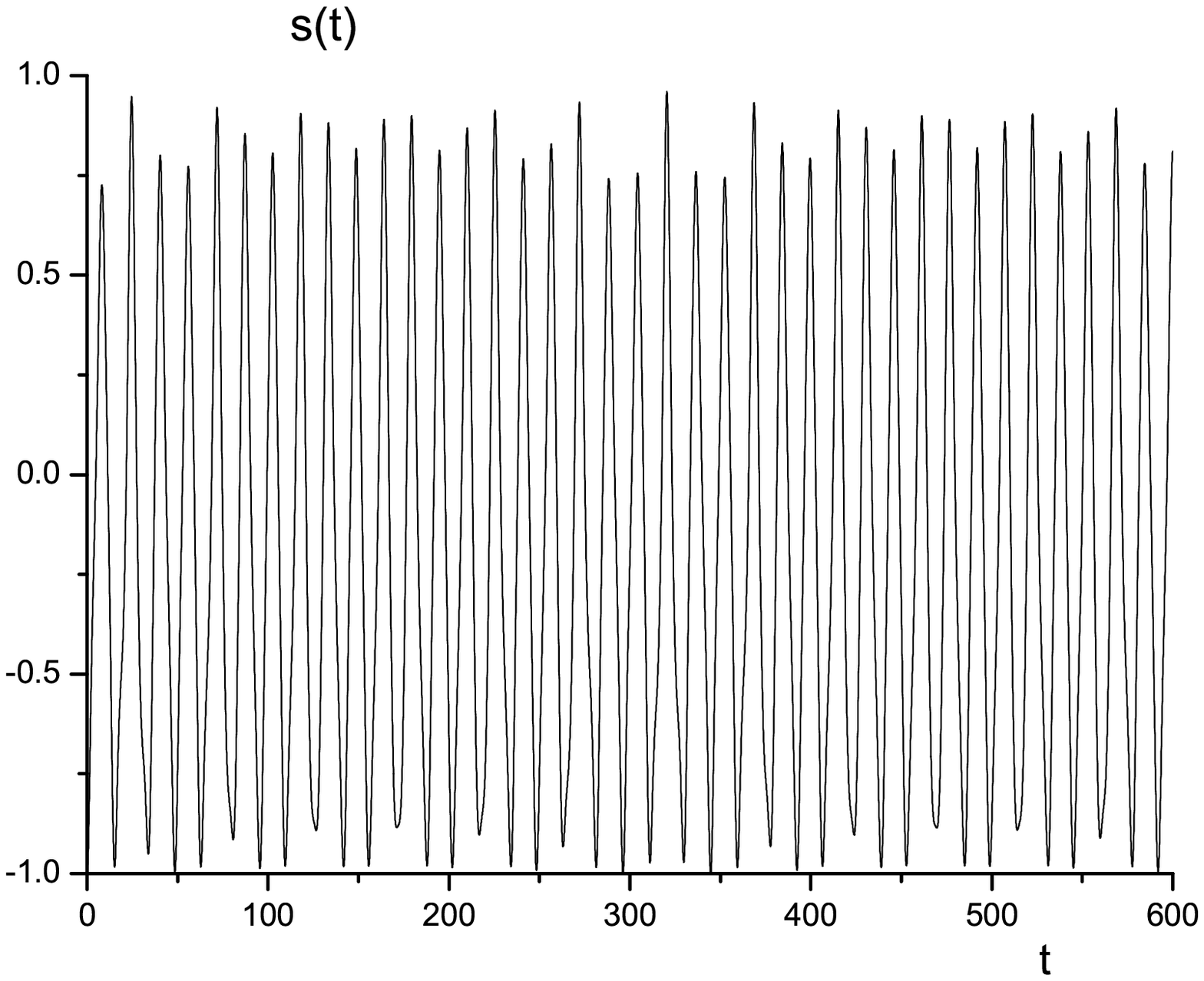}
b) \includegraphics[width=7cm]{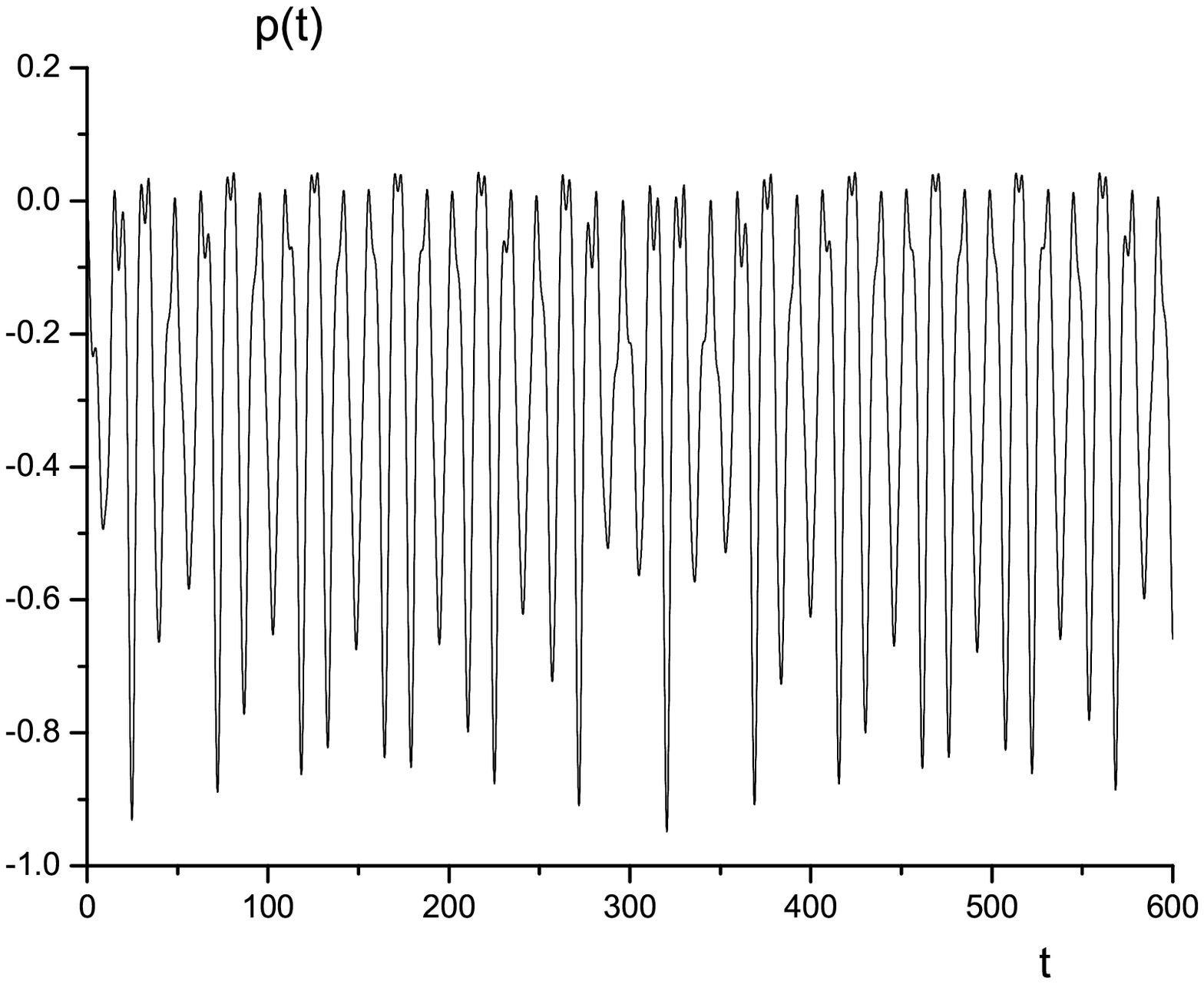}
}
\caption{\label{timeEvolSlavaFig2} Mode unlocked regime for the three-mode
case. All parameters and initial conditions are the same as
in Fig. 1, except $b_{ij}=0.55\alpha$. Again, time is shown in
units of $\alpha^{-1}$. The population differences: (a) $s(t)$;
(b) $p(t)$. }
\end{figure}

We have numerically investigated the behaviour of solutions to Eqs. (\ref{org45})
for various parameters $\alpha_i$, $b_{ij}$, and $\delta_i$. The latter parameter,
playing the role of an effective detuning, was assumed to be small,
$\delta_i\ll 1$. Different initial conditions have been considered in the
range $-1\leq s_0\leq 1$, $-1\leq p_0\leq 1$, $0\leq x_0\leq 2\pi$,
$0\leq y_0\leq 2\pi$. For small $b_{ij}\ll\alpha_i$, the solutions for the
population differences $s$ and $p$ demonstrate a kind of nonlinear Rabi
oscillations in the mode locked regime, when $s(t)$ and $p(t)$ do not cross
the zero line, being always either above or below it, depending on initial
conditions. This mode locked regime is the same as in the case of two
modes, studied earlier
\cite{orgcit8,orgcit14,orgcit17,orgcit21,orgcit22}.
The difference with the two-mode
case is that the Rabi-type oscillations look slightly more complicated, being
quasiperiodic but not periodic. Increasing $b_{ij}$ results in the increase
of the oscillation amplitudes, and, after a critical value of $b_{ij}$, the
mode unlocked regime arises, when either $s(t)$, or $p(t)$, or both of them,
oscillate in the whole interval [-1,1]. To illustrate these two regimes,
we present numerical calculations accomplished for equal parameters
$\alpha_{ij}=\alpha$ with the initial conditions $s_0=-1$, $p_0=0$, $x_0=y_0=0$.
Figure \ref{timeEvolSlavaFig1}
shows the mode locked regime, when $s(t)<0$ for all $t\geq 0$, and
$p(t)$ is also negative for almost all $t$.
Figure \ref{timeEvolSlavaFig2} demonstrates the mode
unlocked regime, when $s(t)\in[-1,1]$, while $p(t)$ is yet almost always
negative. Increasing further $b_{ij}$ leads to the situation, when $p(t)$
starts also oscillating in the interval [-1,1]. However, at large $b_{ij}$,
the temporal behaviour of solutions becomes quite unstable resembling
chaotic motion. To better understand what happens, it is necessary to
study the phase portrait of the dynamical system, that is, one has to
find the fixed points and to analyze their stability. However, before  
turning to this task, we will compare the results of the averaging method  
described above with those of a direct numerical simulation of the GPE for 
several special cases.  
\section{Direct numerical simulation of time evolution}
\label{sec:numsim}
Although
the averaging technique is a well grounded mathematical method
\cite{orgcit43,orgcit47}, it is interesting to show that
the main effects, described above, also appear in a direct numerical
simulation of the time evolution described by Eq. (\ref{org1}).
One important reason is that the results of the averaging
procedure are basically independent of the details of the
trapping potential, while a
direct numerical simulation is affected by it. We consider
two cases: in an anharmonic potential the conditions for the
validity of Eq.~(\ref{org36}) can well be met and reasonable
agreement can be expected. On the other hand, in a harmonic trap
the averaging method may break down. We will demonstrate
both situations using the numerical methods and parameter
settings which are described in the Appendix.

\subsection{Anharmonic trap with linear driving field}

\begin{figure}
\centerline{
a) \includegraphics[width=7cm]{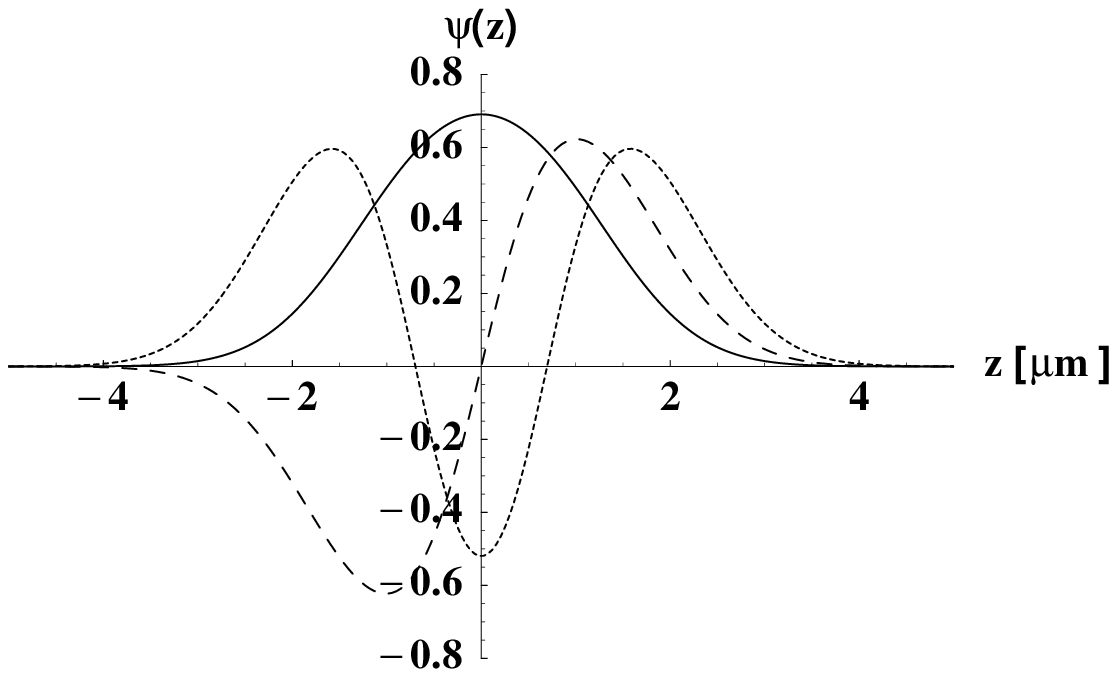}
b) \includegraphics[width=7cm]{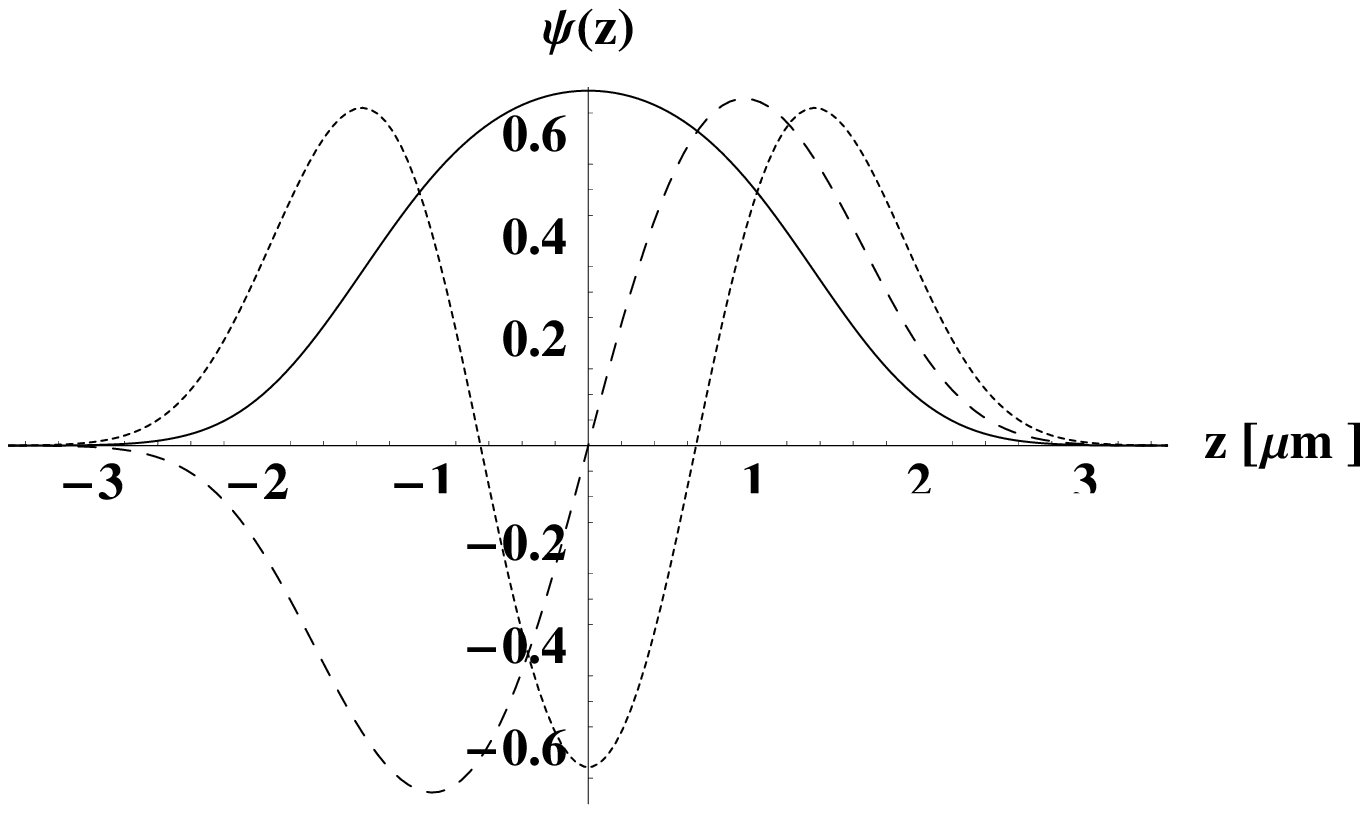}
}
\caption{\label{fig-modes} The three nonlinear coherent modes
of lowest energy for 1000 $^{87}$Rb atoms in two different
trapping potentials. The solid line corresponds to the ground
state, the dashed (dotted) line to the first (second) excited
mode, respectively.
a) harmonic trap with frequency $\omega_z = 600$/s,
b) anharmonic trap of the form $U_0 z^4$ with 
$U_0 = 10^{-32}$ J/$\mu$m$^4$.}
\end{figure}

The case of an anharmonic trapping potential is most suited
to demonstrate the benefits of the few-mode averaging method.
We considered a BEC moving in a potential of
the form $U_0 z^4$
with $U_0 = 10^{-32}$ J/$\mu$m$^4$ and a linear driving potential
of the form $V_1 = \beta_1 z$  (in the notation of Eq.~(\ref{org6}))
which was tuned to resonance with the $\varphi_0 \rightarrow \varphi_1$
transition so that $\omega_1 = (E_1 -E_0)/\hbar \approx 610$/s. 
In absence of the driving potential the BEC can occupy several
stationary nonlinear coherent modes, three of which are displayed
in Fig.~\ref{fig-modes}. The respective parameters appearing
in Eq.~(\ref{org6}) are found to be $\alpha_{12} = 121.7 s^{-1}$,
$\alpha_{13} = 46.3 s^{-1}$, $\alpha_{21} = 144.2 s^{-1}$,
$\alpha_{23} = 88.9 s^{-1}$, $\alpha_{31} = 91.8 s^{-1}$,
and $\alpha_{32} = 111.9 s^{-1}$.

\begin{figure}
\centerline{
a) \includegraphics[width=7cm]{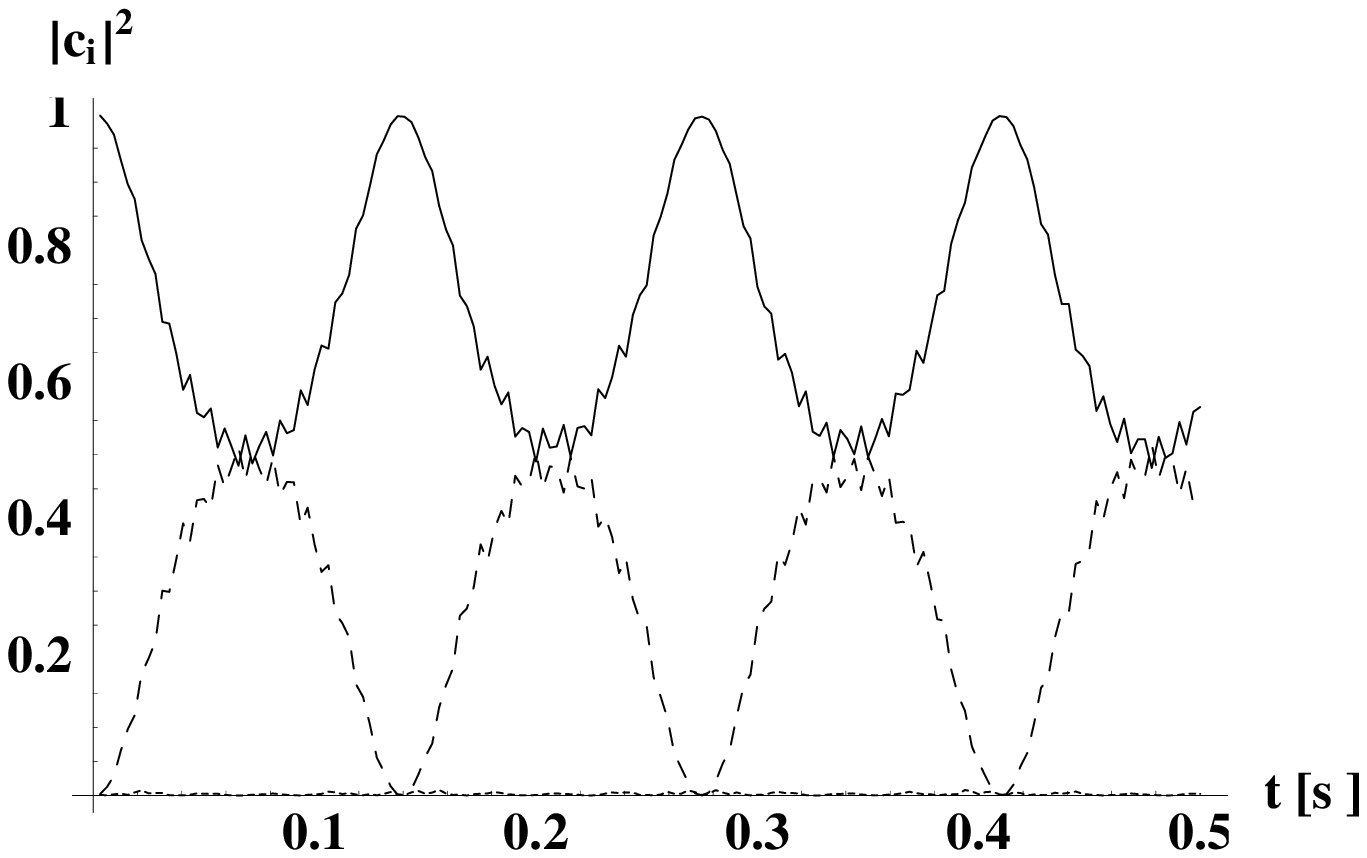}
b) \includegraphics[width=7cm]{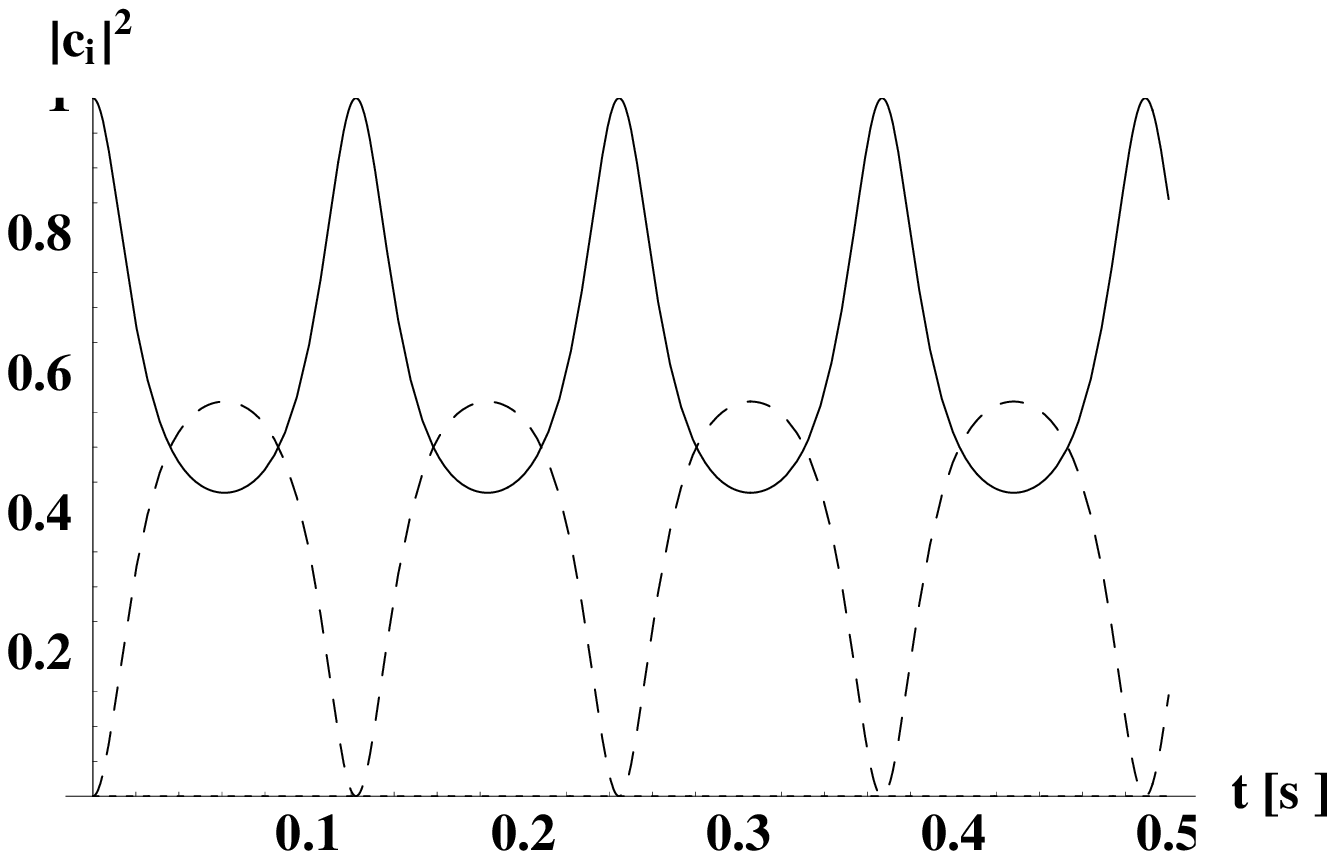}
}
\caption{\label{fig-anharmPotSubcrit} Mode coefficients averaged over
a time of 3 ms
for a BEC in an anharmonic trap $\propto z^4$
driven resonantly by a slightly subcritical linear potential. a) 
results of direct numerical simulation,
b) predictions of the averaging method (see text for
more details). Solid line: $|c_0(t)|^2$, dashed line: $|c_1(t)|^2$,
dotted line: $|c_2(t)|^2$. }
\end{figure}

\begin{figure}
\centerline{
a) \includegraphics[width=7cm]{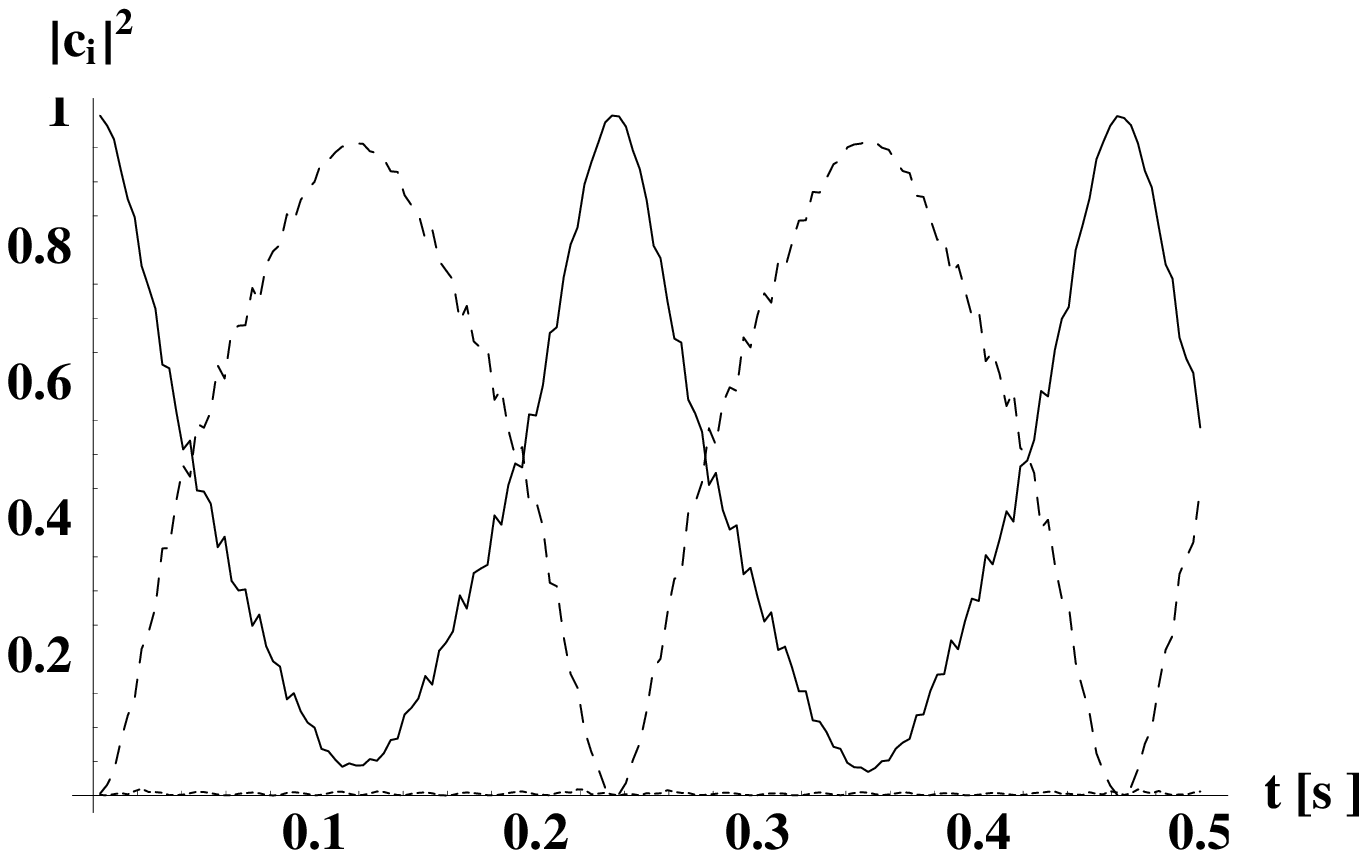}
b) \includegraphics[width=7cm]{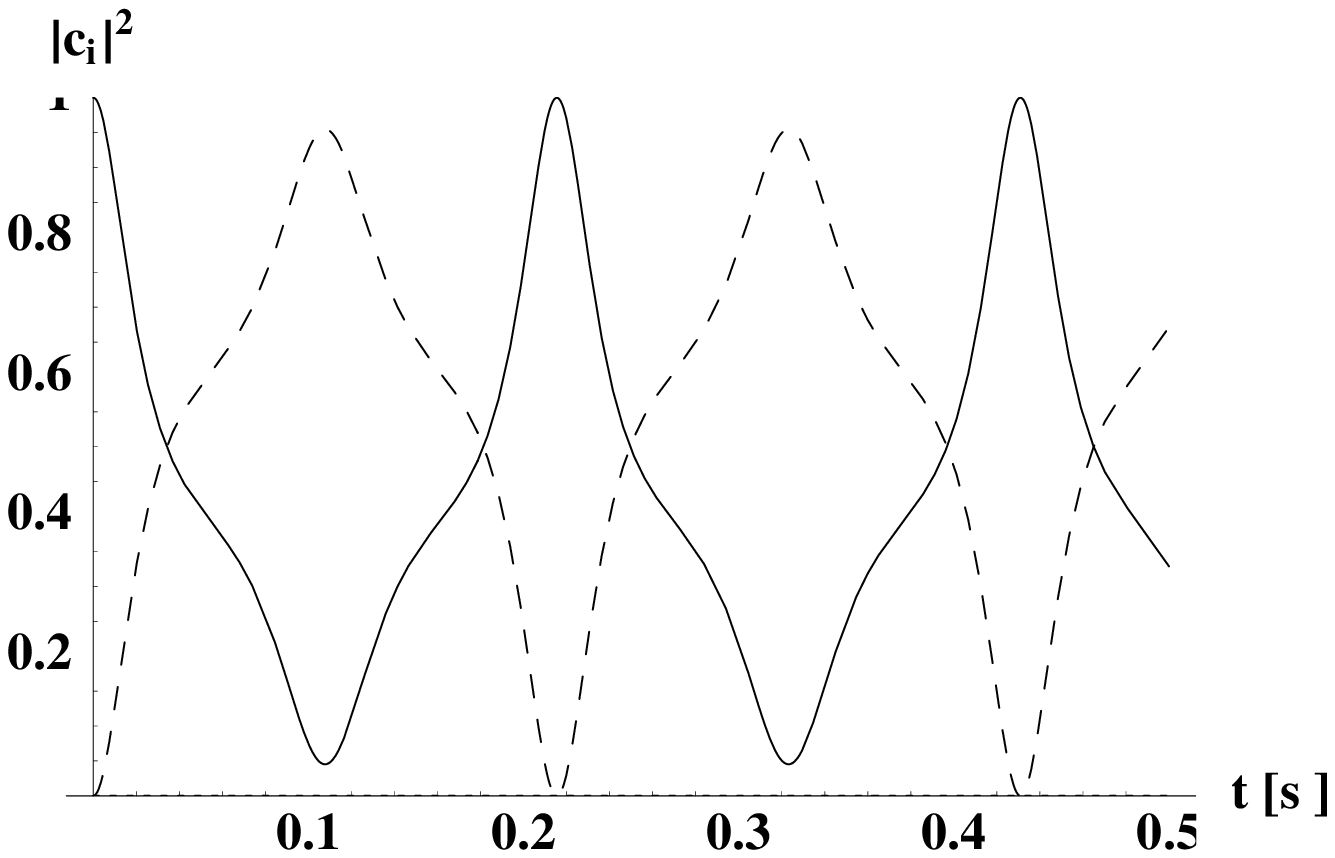}
}
\caption{\label{fig-anharmPotSupercrit} The same as 
Fig.~\ref{fig-anharmPotSubcrit}, but for a slightly supercritical
driving force.}
\end{figure}

As predicted by the averaging method our simulation showed a critical 
behaviour: For $\beta_1$ below a certain value $\beta_c \approx 7.3
\times 10^{-33}$ J/$\mu$m the population transferred to the first excited
state is bounded to be smaller than about 0.5. An example for this
behaviour can be seen in Fig.~\ref{fig-anharmPotSubcrit} a) which corresponds
to a slightly subcritical value of $\beta_1 = 6.9 \times 10^{-33}$ J/$\mu$m.
For $\beta_1 > \beta_c$ this upper bound suddenly disappears and the
population of the first excited states can take almost all values
between 0 and 1. This can be seen in Fig.~\ref{fig-anharmPotSupercrit} a)
which displays the time evolution for a slightly supercritical
driving field with $\beta_1 = 7.7 \times 10^{-33}$ J/$\mu$m. 

The corresponding predictions of the averaging technique can be seen
in Figs~~\ref{fig-anharmPotSubcrit} b) and
\ref{fig-anharmPotSupercrit} b), respectively. Since the period and  
amplitude of the populations $|c_i|^2$ are strongly varying
around the critical value of $\beta_1$, we have chosen the values
$\beta_1 = 1.0\times 10^{-32}$ J/$\mu$m for
\ref{fig-anharmPotSubcrit} b) and
$\beta_1 = 1.02\times 10^{-32}$ J/$\mu$m for
\ref{fig-anharmPotSupercrit} b),
which are close to the critical value as predicted
by the averaging method.
The found critical values of $\beta_c$ are in reasonable agreement, 
the difference being of about
$25\%$, which is, actually, the accuracy of the
averaging technique for the given set of parameters.

\subsection{Harmonic trap with anharmonic driving field}

This case serves as an example of how the breakdown of the
conditions (\ref{org13}) and (\ref{org15}) results in wrong
predictions of the mode expansion method. The case
of a harmonic trap is a very special case with this
respect: if, in absence of interaction, the transition between
two neighbouring modes is resonantly driven, then
this is also the case for a transition between
any other neighbouring modes. It therefore is never possible
to  consider only a small number of states because other
states are quickly populated as well. Only if one is
sufficiently far away from resonance a few-mode model
can be expected to work well, but then the transition rate
between those modes is also low. These conclusions remain
qualitatively also valid in the presence of interaction.

To demonstrate the breakdown of the mode expansion method
we consider a harmonic
potential of the form $U_z(z) = m_0 \omega_z^2 z^2 /2$ with
$\omega_z = 600$/s, and a screened cubic driving field
which, in the notation of Eq.~(\ref{org6}), is given by
$V_i(z) = \beta_i z^3 \exp (-z^2 /w^2)$, $i=1,2$,
with $w=4.9 \mu$m. We introduced the exponential screening
since a purely cubic driving field plus a harmonic potential
is unbounded from below. A linear driving field cannot be
used because, according to the theorem given above,
only leads to an oscillation of the BEC without changing
its shape. The two driving frequencies are chosen to be
at resonance with the $\varphi_0 \rightarrow \varphi_1$
transition, $\omega_1 = (E_1 -E_0)/\hbar \approx 536$/s,
and the $\varphi_1 \rightarrow \varphi_2$
transition, $\omega_2 = (E_2 -E_1)/\hbar \approx 568$/s.

\begin{figure}
\centerline{
a) \includegraphics[width=7cm]{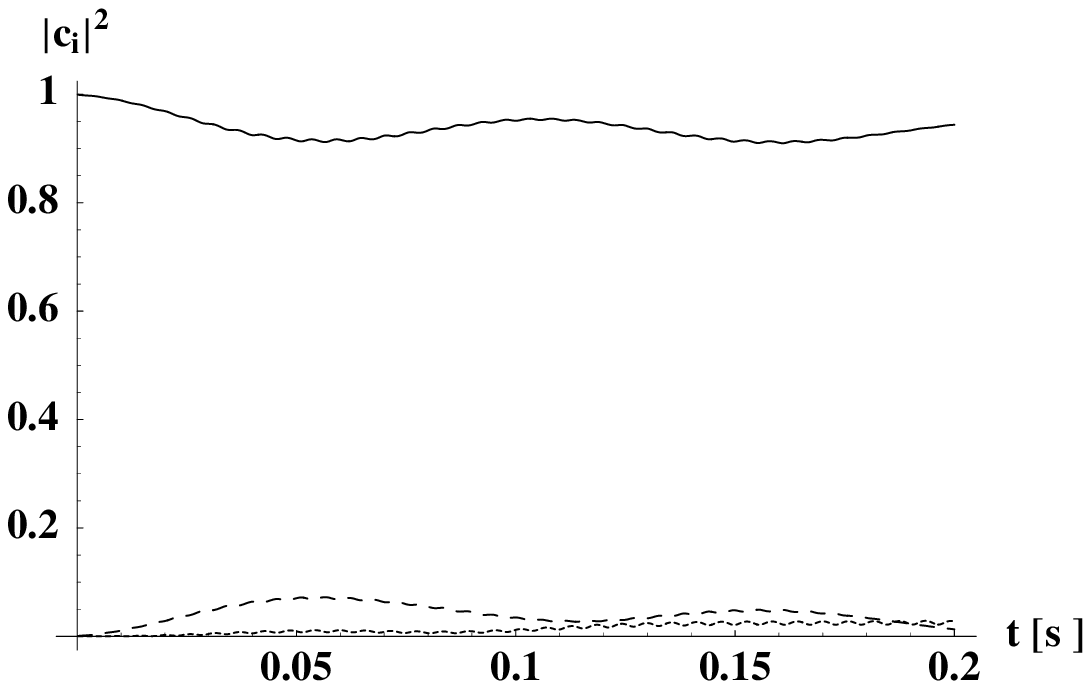}
b) \includegraphics[width=7cm]{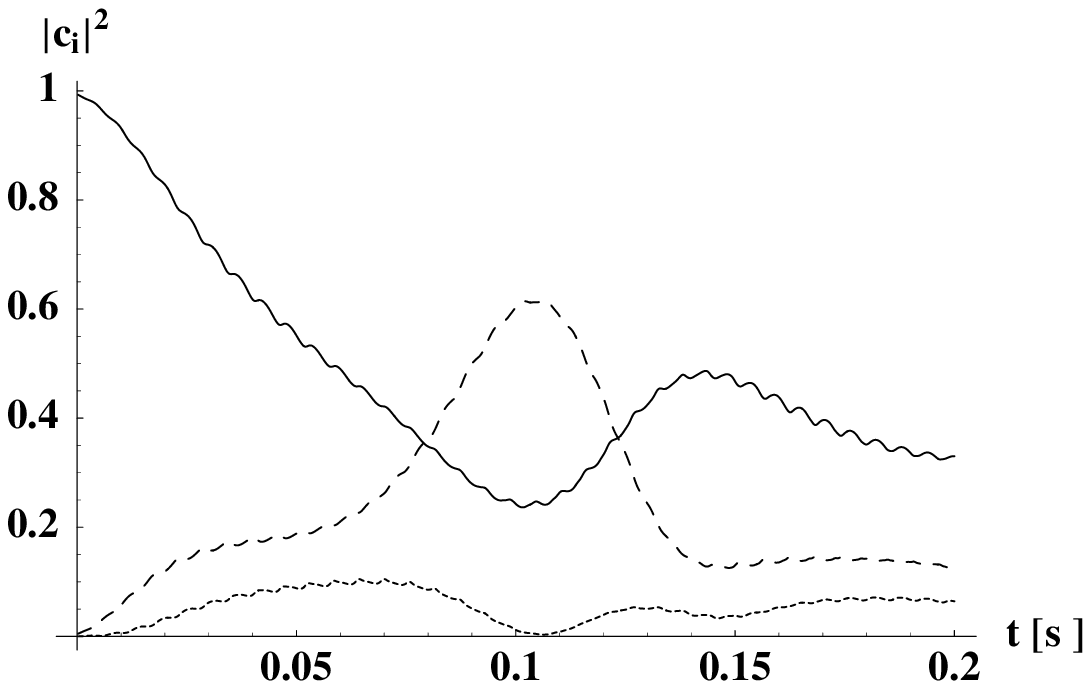}
}
\caption{\label{fig-cubicScreened} Mode coefficients averaged over
a time of 5.3 ms
for a BEC in a harmonic trap driven resonantly by a screened cubic
potential. a) weak driving force with
$\beta_1=0.7869 \times 10^{-33}$ J/$\mu$m$^3$
and $\beta_2=0.344\times 10^{-33}$ J/$\mu$m$^3$.
b) strong driving force with $\beta_1=1.96734\times 10^{-33} $ J/$\mu$m$^3$
and $\beta_2=0.859975\times 10^{-33}$ J/$\mu$m$^3$.
Solid line: $|c_0(t)|^2$, dashed line: $|c_1(t)|^2$,
dotted line: $|c_2(t)|^2$.}
\end{figure}

The relative strength of the two driving fields was chosen
to achieve  $(\varphi_1 ,V_1 \varphi_0) =(\varphi_2 ,V_2 \varphi_1)$
by setting the appropriate values for $\beta_1$ and $\beta_2$.
The results are shown in Fig.~\ref{fig-cubicScreened}.
For weak driving fields, $\beta_1=0.7869 \times 10^{-33}$ J/$\mu$m$^3$
and $\beta_2=0.344\times 10^{-33}$ J/$\mu$m$^3$,
one achieves reasonable agreement with the predictions
of the averaging procedure, see Fig.~\ref{fig-cubicScreened}a).
For a strong driving force with $\beta_1=1.96734\times 10^{-33} $ J/$\mu$m$^3$
and $\beta_2=0.859975\times 10^{-33}$ J/$\mu$m$^3$
the excitation of higher modes becomes significant, which
is not surprising because the driving field provides more energy
which can also excite higher levels. This behaviour can
be seen in Fig.~\ref{fig-cubicScreened}b) which also indicates
that the time evolution deviates strongly from the predictions
of the averaging method.

\section{Stationary Solutions and Stability Analysis}
\label{sec:stabil}
To find out the stationary solutions for the dynamics of the three-mode case
and to analyse the stability of these solutions, it is convenient to work
with the variables
\begin{equation}
\label{org46}
f_j \equiv |c_j| \qquad (j=1,2,3)
\end{equation}
and the relative phases (\ref{org44}). The fractional mode populations are  
expressed through the amplitudes (\ref{org46}) as $n_j=f_j^2$. 

Using the  variables $f_j$, $x$, and $y$ 
one can rewrite Eqs. (\ref{org36}) for the cascade generation
in the form

$$
\frac{df_1}{dt} = \frac{1}{2}\; b_{12}\; f_2\; \sin x \; ,
$$
$$
\frac{df_2}{dt} = -\; \frac{1}{2}\; b_{12}\; f_1\; \sin x +
\frac{1}{2}\; b_{23}\; f_3\; \sin y \; ,
$$
$$
\frac{df_3}{dt} = -\; \frac{1}{2}\; b_{23}\; f_2\; \sin y\; ,
$$
$$
\frac{dx}{dt} = \Delta_{21} - \alpha_{21}\; f_1^2 + \alpha_{12}\; f_2^2
+ ( \alpha_{13} - \alpha_{23} )\; f_3^2 + b_{12}\;
\frac{f_2^2-f_1^2}{2f_1f_2}\; \cos x - b_{23}\;
\frac{f_3}{2f_2}\; \cos y \; ,
$$
\begin{equation}
\label{org47}
\frac{dy}{dt} = \Delta_{32}  + ( \alpha_{21} - \alpha_{31} )\; f_1^2 -
\alpha_{32}\; f_2^2 + \alpha_{23}\; f_3^2  +
b_{12}\;  \frac{f_1}{2f_2}\; \cos x +
b_{23}\; \frac{f_3^2-f_2^2}{2f_2f_3}\; \cos y \; .
\end{equation}
There are here yet too many parameters, because of which the analysis of
Eqs. (\ref{org47}) is yet too complicated. To simplify the consideration,
we may involve a realistic approximation, when there is no detuning, the
amplitudes $b_{ij}$, due to the applied alternating fields, are taken to
be the same, and the parameters $\alpha_{ij}$ are close to each other.
That is, we set
\begin{equation}
\label{org48}
\alpha\equiv \alpha_{ij} \; , \qquad b\equiv \frac{b_{ij}}{\alpha} \; ,
\qquad \Delta_{ij} = 0 \; .
\end{equation}
It is also convenient to measure time in units of $\alpha^{-1}$. To return
back to time units, one should replace $t$ by $\alpha t$. Then Eqs. (\ref{org47})
reduce to
 $$
\frac{df_1}{dt} = \frac{b}{2}\; f_2 \; \sin x \; ,
$$
$$
\frac{df_2}{dt} = -\; \frac{b}{2}\; f_1 \; \sin x +
\frac{b}{2}\; f_3 \; \sin y \; ,
$$
$$
\frac{df_3}{dt} = -\; \frac{b}{2}\; f_2 \; \sin y \; ,
$$
$$
\frac{dx}{dt} = f_2^2 - f_1^2 + b\;
\frac{f_2^2 - f_1^2}{2f_1f_2} \; \cos x -
b \; \frac{f_3}{2f_2}\; \cos y \; ,
$$
\begin{equation}
\label{org49}
\frac{dy}{dt} = f_3^2 - f_2^2 + b\; \frac{f_1}{2f_2}\; \cos x
+ b\; \frac{f_3^2 - f_2^2}{2f_2f_3} \; \cos y \; .
\end{equation}
In addition, because of Eqs. (\ref{org40}) and (\ref{org46}), there is the relation
\begin{equation}
\label{org50}
f_1^2 + f_2^2 + f_3^2 =  1 \; .
\end{equation}
This relation is automatically supported by Eqs. (\ref{org49}) as well as
by Eqs. (\ref{org47}), provided it is valid for the initial $f_i(0)$.
Equations (\ref{org49}) are much easier to analyse than Eqs. (\ref{org47}). At the
same time, the mathematical structure of these equations is similar,
so the behaviour of their solutions should be close to each other.

In Eqs. (\ref{org49}), there is the sole dimensionless parameter $b$, defined
in Eq. (\ref{org48}). Since the parameter $\alpha$ can be positive as well as
negative, then $b$ can, generally, be of both signs too. But Eqs.
(\ref{org49}) enjoy a nice symmetry property, being invariant under the
change
\begin{equation}
\label{org51}
b\rightarrow - b \; , \qquad x\rightarrow x\pm \pi\; , \qquad y\rightarrow y\pm \pi \; .
\end{equation}
Therefore, in what follows, it is sufficient to consider only the
case of positive $b\geq 0$.

There exists a special solution of Eqs. (\ref{org49}), for which
\begin{equation}
\label{org52}
f_1 = f_3 \; , \qquad x = - y
\end{equation}
for all times $t\geq 0$. Then the problem reduces to a kind of a
two-mode case described by the equations
$$
\frac{df_1}{dt} = \frac{b}{2}\; f_2\; \sin x \; ,
$$
$$
\frac{df_2}{dt} = -b\; f_1\; \sin x \; ,
$$
$$
\frac{dx}{dt} = f_2^2 - f_1^2 +
b\; \frac{2f_2^2-1}{2f_1f_2}\; \cos x \; .
$$
This reduction, however, becomes possible as a result of the
approximation when all interaction amplitudes $\alpha_{ij}=\alpha$ are
assumed to be equal. More precisely, the minimal requirements
necessary for the existence of solution (\ref{org52}) are
\begin{equation}
\label{org53}
\alpha_{12}=\alpha_{32} \; , \qquad \alpha_{13}=\alpha_{31} \; , \qquad
b_{12}=b_{23}\; , \qquad \Delta_{21}=\Delta_{32} \; .
\end{equation}
It is feasible, in principle, to choose such modulating fields
and a trapping potential that Eqs. (\ref{org53}) be valid. In addition, the
initial conditions are to be such that $f_1(0)=f_3(0)$.

The stationary solutions of Eqs. (\ref{org49}) are obtained by equating to
zero their right-hand sides, keeping in mind that $b\neq 0$ and
$f_2$ is not identically zero. The corresponding fixed-point
equations for the phase differences are
\begin{equation}
\label{org54}
\sin x= \sin y = 0 \; , \qquad \cos x = \pm 1 \; , \qquad
\cos y =\pm 1 \; .
\end{equation}
The equations for the amplitudes are rather cumbersome, and we
shall not write them down. We shall solve these equations
numerically, calculating the fixed-point amplitudes $f_i^*=f_i^*(b)$
as functions of the pumping parameter $b$ and then finding the
related stationary solutions
\begin{equation}
\label{org55}
n_i^*(b) \equiv |f_i^*(b)|^2
\end{equation}
for the fractional mode populations. Simultaneously, we accomplish
the stability analysis by calculating the Jacobian matrix for Eqs.
(\ref{org49}), evaluated at the related fixed points. The eigenvalues of
this matrix define the characteristic exponents characterizing the
type of stability \cite{orgcit46}.

\begin{figure}
\centerline{
\includegraphics[width=9cm]{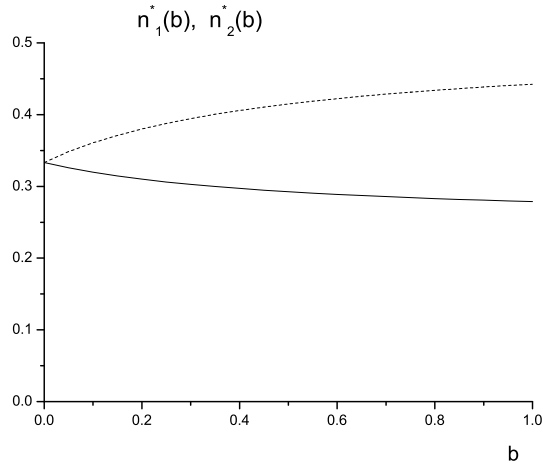}
}
\caption{\label{stabilityFig1} Stationary solutions for the case $x^*=y^*=0$
and $n_1^*=n_3^*$ as functions of the pumping parameter $b$:
stable $n_1^*(b)=n_3^*(b)$ (solid line), stable $n_2^*(b)$
(dashed line).}
\end{figure}

The first stationary solution is given by $x^*=y^*=0$ and $n_1^*=n_3^*$. The
related mode populations (\ref{org55}) are shown in
Fig. \ref{stabilityFig1}. This fixed point is
neutrally stable, being a center with the characteristic exponents $\Lambda_1=0$
and imaginary $\Lambda_2=\Lambda_3^*$, $\Lambda_4=\Lambda_5^*$. When $b$ varies in the
interval $0.1\leq b\leq 1$, the absolute values of $\Lambda_{2,3}$ and
$\Lambda_{4,5}$ change in the range $0.192\leq|\Lambda_{2,3}|\leq 0.865$ and
$0.351\leq|\Lambda_{4,5}| \leq 1.755$.

\begin{figure}
\centerline{
a) \includegraphics[width=7cm]{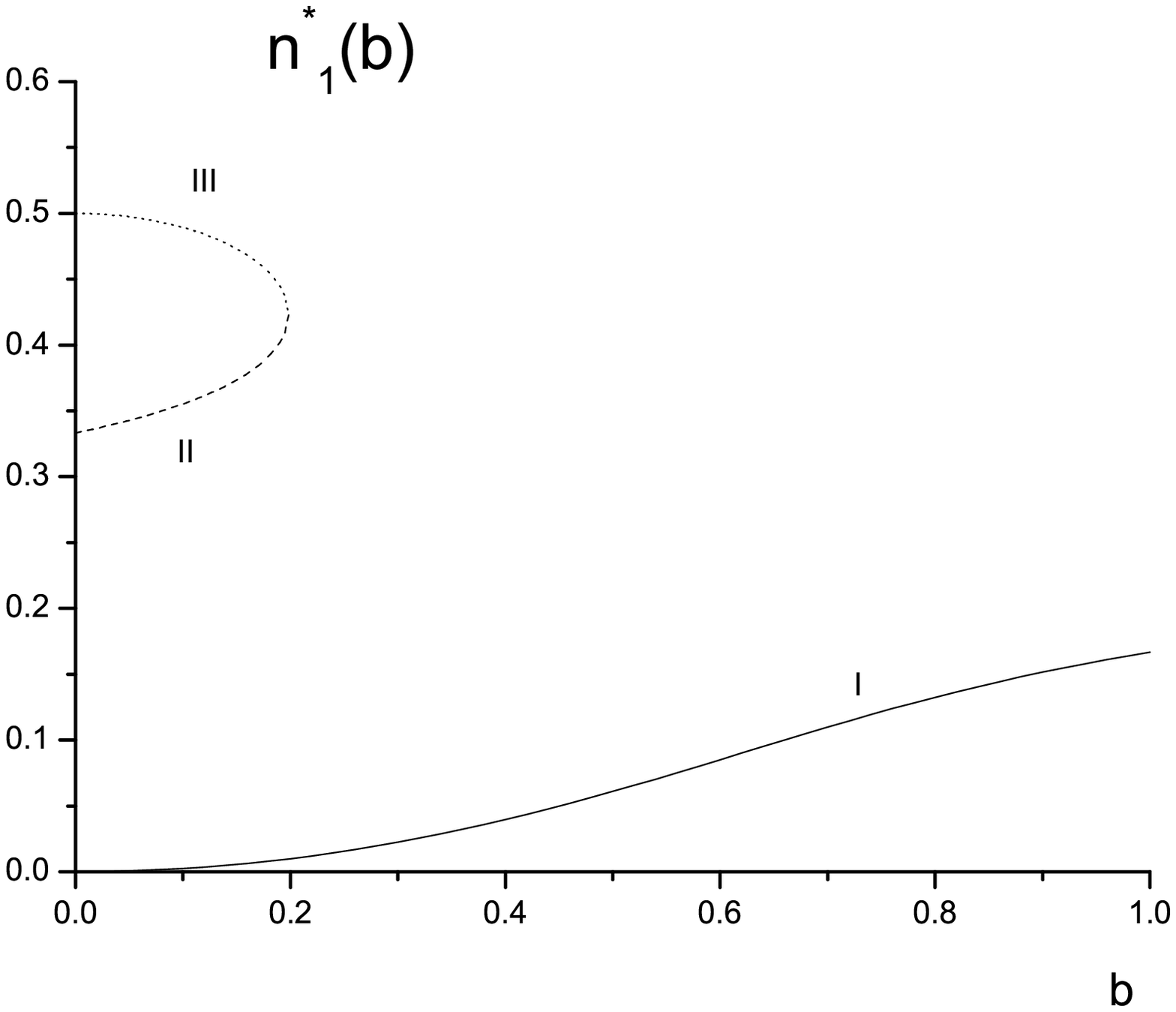}
b) \includegraphics[width=7cm]{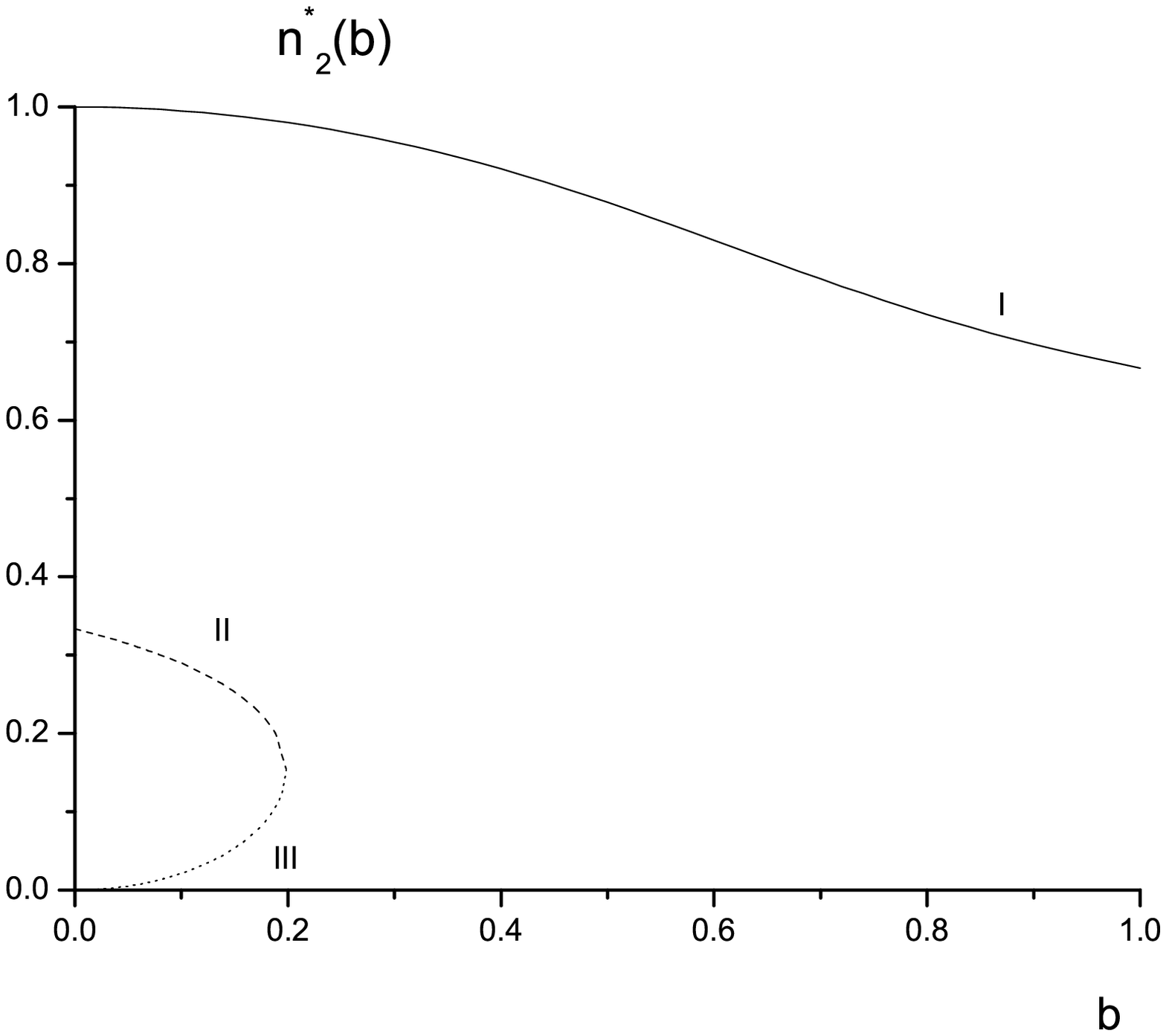}
}
\caption{\label{stabilityFig2} Stationary solutions related to the fixed
point $x^*=y^*=\pi$ and $n_1^*=n_3^*$ as functions of the
pumping parameter $b$. Stable branch I (solid line), unstable
branch II (dashed line), and unstable branch III (dashed
line): (a) $n_1^*(b)=n_3^*(b)$; (b) $n_2^*(b)$.}
\end{figure}

The second fixed point is defined by $x^*=y^*=\pi$ and the branch I in
Fig. \ref{stabilityFig2}a) for $n_1^*=n_3^*$ and in
Fig. \ref{stabilityFig2}b) for $n_2^*$. This fixed point
is also a center with the characteristic exponents $\Lambda_1=0$ and imaginary
$\Lambda_2=\Lambda_3^*$, $\Lambda_4=\Lambda_5^*$. For $b\in[0,1]$, one has
$|\Lambda_{2,3}|\sim|\Lambda_{4,5}|\sim 1$.

The third fixed point corresponds to $x^*=y^*=\pi$ and the branch II in
Figs. \ref{stabilityFig2}a) and
 \ref{stabilityFig2}b) for $n_1^*=n_3^*$ and $n_2^*$, respectively. The
characteristic exponents are $\Lambda_1=0$ and real $\Lambda_2=-\Lambda_3$,
$\Lambda_4=-\Lambda_5$, which shows that this point is unstable. The absolute
values are $|\Lambda_{2,3}|\sim|\Lambda_{4,5}|\sim 0.1$. The solution exists
for $0\leq b\leq b_0^*$, with $b_0^*=0.198431$.

The fourth fixed point is given by $x^*=y^*=\pi$ and the branch III in
Figs.~\ref{stabilityFig2}a) and
 \ref{stabilityFig2}b) for $n_1^*=n_3^*$ and $n_2^*$. The characteristic
exponents are $\Lambda_1=0$, imaginary $\Lambda_2=\Lambda_3^*$, and real
$\Lambda_4=-\Lambda_5$, with $|\Lambda_{2,3}|\sim|\Lambda_{4,5}|\sim 0.1$. This
solution is unstable and exists under $0\leq b\leq b_0^*$, with
$b_0^*=0.198431$.

\begin{figure}
\centerline{
a) \includegraphics[width=5cm]{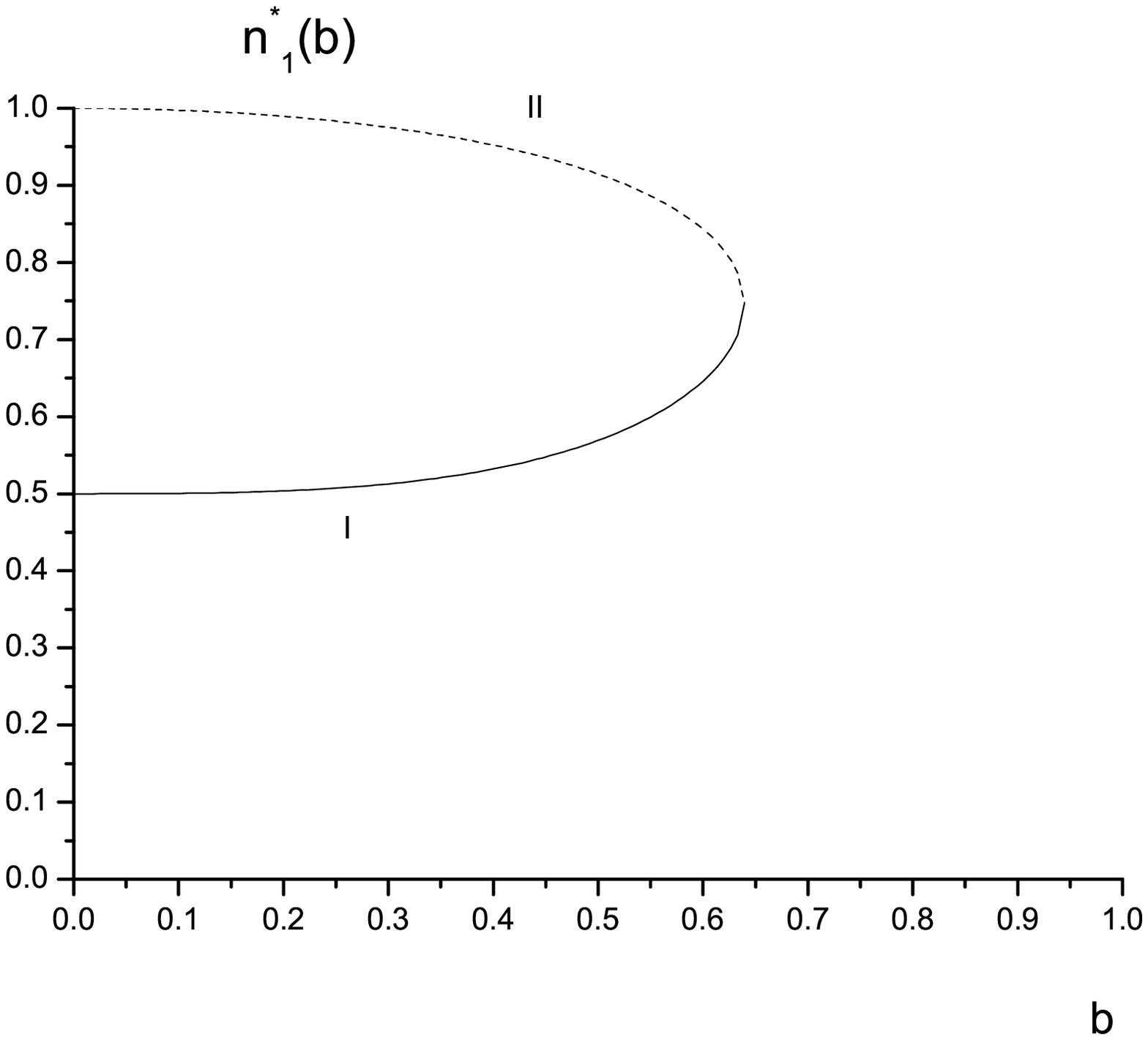}
b) \includegraphics[width=5cm]{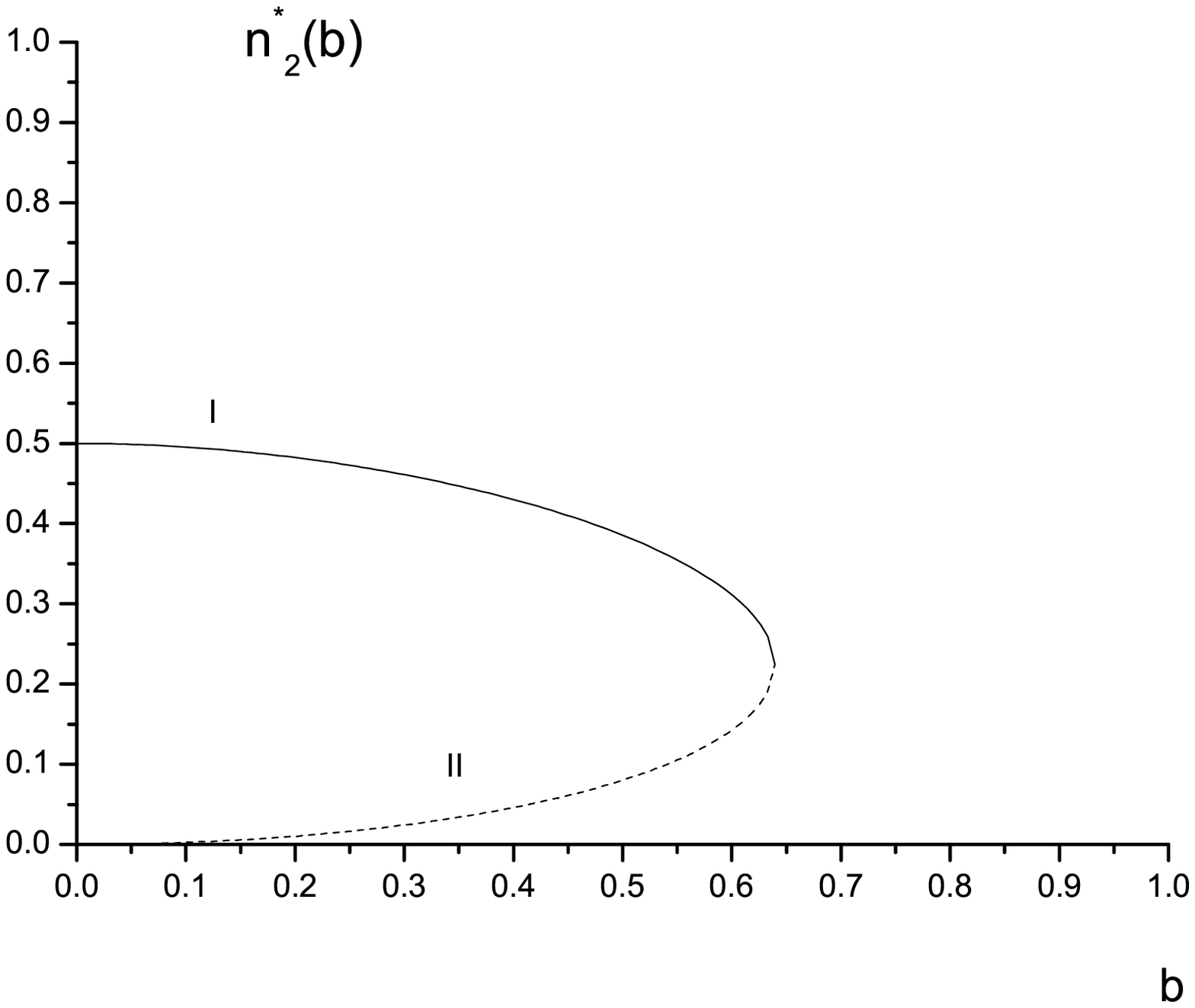}
c) \includegraphics[width=5cm]{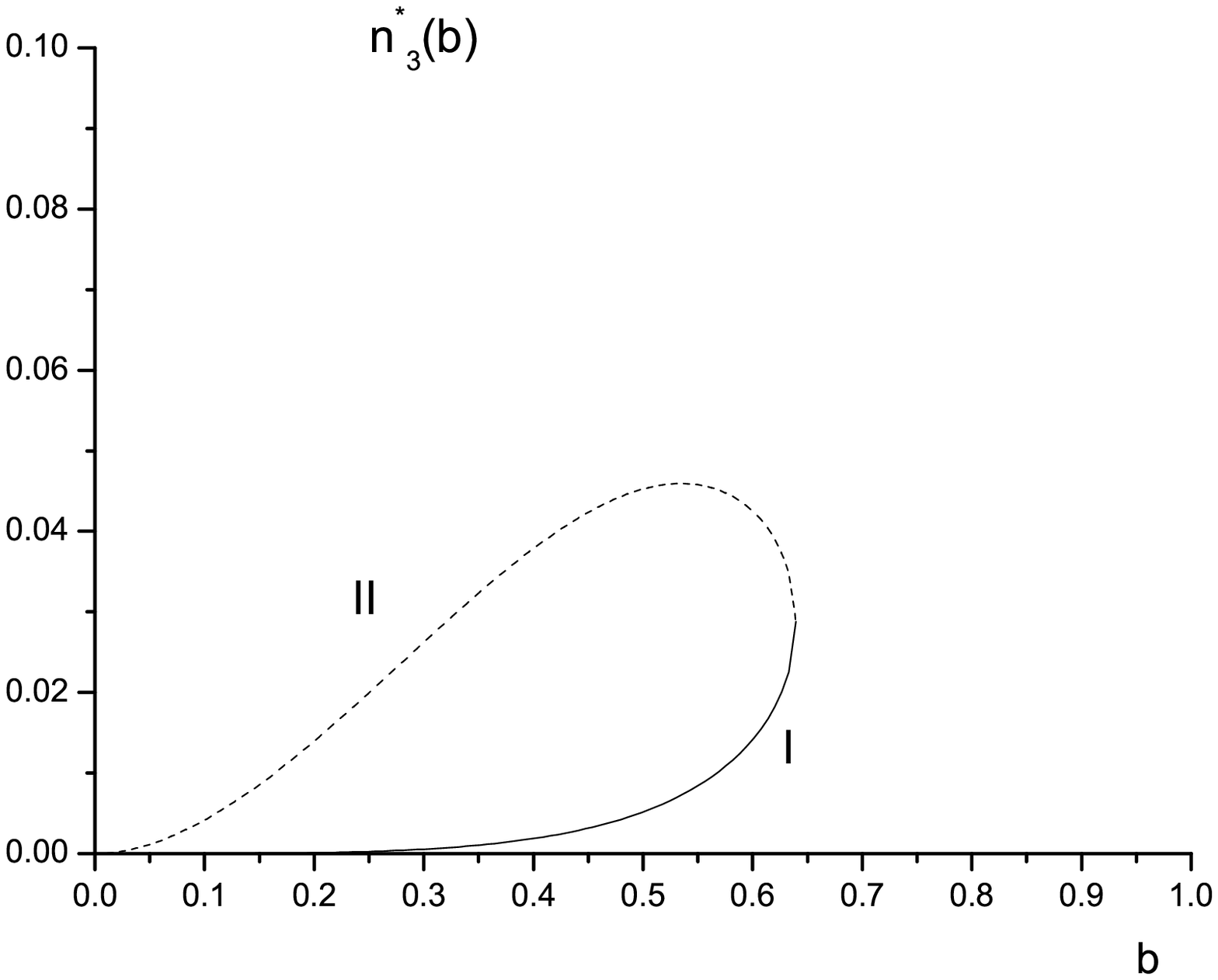}
}
\caption{\label{stabilityFig3} Stationary solutions associated with the fixed
point $x^*=y^*=\pi$ and $n_1^*>n_3^*$ as functions of $b$.
Stable branch I (solid line) and unstable branch II (dashed
line): (a) $n_1^*(b)$; (b) $n_2^*(b)$; (c) $n_3^*(b)$.}
\end{figure}

The fifth fixed point is described by $x^*=y^*=\pi$ and the branch I in
Fig.~\ref{stabilityFig3} for $n_1^*(b)$, $n_2(b)$, and $n_3^*(b)$,
respectively. The characteristic exponents are
$\Lambda_1=0$ and imaginary $\Lambda_2=\Lambda_3^*$,
$\Lambda_4=\Lambda_5^*$, which shows that this point is a center. The solution
exists for $b$ in the range $0\leq b\leq b_c^*$, with $b_c^*=0.639448$, where
$|\Lambda_{2,3}|\sim 1$ and $|\Lambda_{4,5}|\leq 0.967$.

The sixth fixed point corresponds to $x^*=y^*=\pi$ and the branch II in
Fig.~\ref{stabilityFig3}
for $n_1^*(b)$, $n_2^*(b)$, and $n_3^*(b)$. The related characteristic
exponents are $\Lambda_1=0$, imaginary $\Lambda_2=\Lambda_3^*$, and real
$\Lambda_4=-\Lambda_5$. This tells that this point is unstable. The solution
exists for $0\leq b\leq b_c^*$, with $b_c^*=0.639448$, where
$|\Lambda_{2,3}|\leq 0.896$ and $|\Lambda_{4,5}|\leq 0.457$.

The stationary solutions for the case $n_1^*\neq n_3^*$ are, actually,
invariant under the interchange of the indices 1 and 3. For concreteness,
we accept that $n_1^*>n_3^*$, as is shown in
Fig.~\ref{stabilityFig3}. Formally, there is
also the seventh fixed point, for which $x^*=y^*=2\pi$ and $n_1^*=n_3^*$.
But this is just the first point shifted in $x^*$ and $y^*$ by $2\pi$.

Selecting from all available fixed points only those that corresponds to
stable stationary solutions, we have three such cases: the point $x^*=y^*=0$,
$n_1^*=n_3^*$, depicted in
Fig.~\ref{stabilityFig1}; the point $x^*=y^*=\pi$, $n_1^*=n_3^*$, shown in
Fig.~\ref{stabilityFig2} as the branch I;
and the point $x^*=y^*=\pi$, $n_1^*>n_3^*$, presented as the branch I in
Fig.~\ref{stabilityFig3}.

Recall that these three stable fixed points exist under the validity of
conditions (\ref{org53}). In reality, it can be quite difficult to satisfy these
conditions exactly. But if these conditions are not valid, then the sole
stable point that remains is the point $x^*=y^*=\pi$, $n_1^*>n_3^*$, which
is shown as the branch I in
Fig.~\ref{stabilityFig3}. This stationary solution exists only
for $b\leq b_c^*$. For larger pumping parameters $b>b_c^*$, there are no
stable (or neutrally stable) fixed points. Hence, the motion for such
large $b$ will be chaotic.

\section{Harmonic Generation and Parametric Conversion}
\label{sec:harmgen}
Employing the presentation (\ref{org14}) in the GPE (\ref{org1}), and
involving the averaging technique \cite{orgcit43}, we come to the equations for
$c_n(t)$, like Eqs. (\ref{org36}). This procedure defines an initial approximation
for $c_n(t)$, which can be called the guiding centers and labelled as
$c_n^{(0)}(t)$. It is possible to obtain corrections to the guiding
centers by using the following steps of the averaging technique
\cite{orgcit43,orgcit47}.
Then we can find the higher approximations $c_n^{(k)}(t)$ describing the
fractional mode populations $|c_n^{(k)}(t)|^2$. To obtain the higher
approximations for $c_n(t)$, we may proceed as follows.

Let us present the solution to the temporal equation (\ref{org1}) as
\begin{equation}
\label{org56}
\varphi({\bf r},t) = \sum_n c_n(t) \; \varphi_n({\bf r},t) \; ,
\end{equation}
where again $c_n(t)$ is a slow function of time, compared to
the fast function $\varphi_n({\bf r},t)$. For any fast function of time
$F(t)$, we define the averaging
\begin{equation}
\label{org57}
<F>\; \equiv \lim_{\tau\rightarrow\infty} \; \frac{1}{\tau} \;
\int_0^\tau \; F(t)\; dt
\end{equation}
over fast oscillations. In particular,
\begin{eqnarray}
\nonumber
<\exp\{ i(\omega_1-\omega_2)t\}>\; = \Delta(\omega_1-\omega_2)
\equiv \left\{ \begin{array}{ll}
1, & \omega_1=\omega_2 \\
0, & \omega_1 \neq \omega_2 \; .
\end{array} \right.
\end{eqnarray}
The functions $\varphi_n({\bf r},t)$ can be taken such that
\begin{equation}
\label{org58}
<\int \varphi_m^*({\bf r},t)\; \varphi_n({\bf r},t)\; d{\bf r}>\; = \delta_{mn} \; .
\end{equation}
Then the amplitudes $c_n(t)$ in Eq. (\ref{org56}) can be obtained as
\begin{equation}
\label{org59}
c_n(t) = \; < \int \varphi_n^*({\bf r},t)\; \varphi({\bf r},t)\; d{\bf r} > \; ,
\end{equation}
and the normalization
\begin{equation}
\label{org60}
\sum_n |c_n(t)|^2 = 1
\end{equation}
is valid. For instance, taking $\varphi_n({\bf r},t)$ in the form
\begin{equation}
\label{org61}
\varphi_n^{(0)}({\bf r},t) = \varphi_n({\bf r}) \;
\exp\left ( - \; \frac{i}{\hbar}\; E_n\; t\right ) \; ,
\end{equation}
we come back to Eq. (\ref{org14}), with all Eqs. (\ref{org58}) to (\ref{org60}) being
evidently satisfied.

Equation (\ref{org59}) can be employed as a relation for an iterative
procedure defined by the rule
\begin{equation}
\label{org62}
c_n^{(k+1)}(t) = \; < \int \varphi_n^{(k)*}({\bf r},t) \;
\varphi^{(k)}({\bf r},t) \; d{\bf r}> \; ,
\end{equation}
where
\begin{equation}
\label{org63}
\varphi^{(k)}({\bf r},t) \equiv \sum_n c_n^{(k)}(t) \;
\varphi_n^{(k)}({\bf r},t) \; .
\end{equation}
Starting from the guiding centers $c_n^{(0)}(t)$ and the form
(\ref{org61}), we get
\begin{equation}
\label{org64}
c_n^{(\ref{org1})}(t) = c_n^{(0)}(t) \; ,
\end{equation}
which follows from the rule (\ref{org62}).

To derive the second-order approximation for $c_n(t)$, we write
\begin{equation}
\label{org65}
\varphi_n^{(\ref{org1})}({\bf r},t) = [ \varphi_n({\bf r}) +
\chi_n({\bf r},t)]\; \exp\left ( -\; \frac{i}{\hbar}\; E_n\; t
\right ) \; .
\end{equation}
Here $\varphi_n({\bf r})$ is a stationary topological mode given by
the eigenproblem (\ref{org4}), and $\chi_n({\bf r},t)$ has to be found
by substituting Eq. (\ref{org65}) into the GPE
(\ref{org1}). In the case of modulating field (\ref{org9}), the
correcting term $\chi_n({\bf r},t)$ can be written as
\begin{equation}
\label{org66}
\chi_n({\bf r},t) = \sum_j \left [ u_{nj}({\bf r}) \;
e^{-i\omega_j t} + v_{nj}^*({\bf r})\; e^{i\omega_j t} \right ] \; ,
\end{equation}
where $j=1,2,\ldots$, and all $\omega_j>0$ can be ordered so that
$0\leq\omega_1\leq\omega_2\leq\ldots$. The functions $u_{nj}({\bf r})$ and
$v_{nj}({\bf r})$ satisfy the equations
$$
\left ( \hat H[\varphi_n] - E_n + N\; A_s\; |\varphi_n|^2 -\hbar\omega_j
\right )\; u_{nj} + N\; A_s\;\varphi_n^2\; v_{nj} =
-\;\frac{1}{2}\; \varphi_n\; B_j^* \; ,
$$
\begin{equation}
\label{org67}
\left ( \hat H[\varphi_n] - E_n + N\; A_s\; |\varphi_n|^2 +\hbar\omega_j
\right )\; v_{nj} + N\; A_s\;(\varphi_n^*)^2\; u_{nj} =
-\;\frac{1}{2}\; \varphi_n^*\; B_j \; ,
\end{equation}
where $\varphi_n=\varphi_n({\bf r})$. Using the first-order form (\ref{org65}) in
Eqs. (\ref{org63}) and (\ref{org62}), we obtain the second-order approximation
$$
c_n^{(\ref{org2})}(t) = c_n^{(0)}(t) + \sum_m c_m^{(0)}(t) \left\{
\sum_i \left [ (\varphi_n,u_{mi}) +(v^*_{ni},\varphi_m)\right ]\;
\Delta(\omega_i -\omega_{nm}) + \right.
$$
$$
\sum_i \left [ (\varphi_n,v_{mi}^*) +
(u_{ni},\varphi_m)\right ] \; \Delta(\omega_i +\omega_{nm}) +
$$
$$
+\sum_{ij}\left [ (u_{ni},u_{mj}) \Delta(\omega_i-\omega_j+\omega_{nm})
+ (u_{ni},v^*_{mj}) \Delta(\omega_i+\omega_j+\omega_{nm}) + \right.
$$
\begin{equation}
\label{org68}
+\left. \left. (v^*_{ni},u_{mj}) \Delta(\omega_i+\omega_j-\omega_{nm})
+ (v^*_{ni},v^*_{mj}) \Delta(\omega_i-\omega_j-\omega_{nm})\right ]
\right\} \; ,
\end{equation}
in which $(u,v)$ denotes the corresponding scalar product. This formula is
valid for an arbitrary number of alternating resonant fields. In the case
of just one resonant field, there should be no summation over the indices
$i$ and $j$ in Eq. (\ref{org68}). Formula (\ref{org68}) shows what are the conditions on the
alternating fields, which induce resonant transitions between topological
modes. These conditions depend on the number of the modulating fields
involved.

Consider the case of two resonantly coupled topological modes,
when
\begin{equation}
\label{org69}
c_n^{(0)}(t) = 0 \qquad (n\neq 1,2) \; ,
\end{equation}
and is nonzero only for $n=1,2$, as follows from the averaging technique
for the guiding centers
\cite{orgcit6,orgcit8}. And suppose that there is the sole
alternating field with a frequency $\omega_1\equiv\omega$. Then the nontrivial
population amplitudes are
$$
c_1^{(\ref{org2})}(t) = c_1^{(0)}(t) + c_2^{(0)}(t) \left [
(\varphi_1,v_2^*) + (u_1,\varphi_2)\right ]\; \Delta(\omega-\omega_{21})+
c_2^{(0)}(t)(u_1,v_2^*)\Delta(2\omega-\omega_{21}) \; ,
$$
\begin{equation}
\label{org70}
c_2^{(\ref{org2})}(t) = c_2^{(0)}(t) + c_1^{(0)}(t) \left [
(\varphi_2,u_1) + (v_2^*,\varphi_1)\right ]\; \Delta(\omega-\omega_{21})+
c_1^{(0)}(t)(v^*_2,u_1)\Delta(2\omega-\omega_{21}) \; ,
\end{equation}
which results from Eq. (\ref{org68}), taking into account that
$\omega_{12}=-\omega_{21}$.

If there are two alternating fields, so that $j=1,2$ in Eq. (\ref{org68}), then
for the nonzero population amplitudes, one has
$$
c_1^{(\ref{org2})} = c_1^{(0)}(t) + c_2^{(0)}\left\{ \left [
(\varphi_1,v_{21}^*)+(u_{11},\varphi_2)\right ] \Delta(\omega_1-\omega_{21}) +
\right.
$$
$$
+ \left [ (\varphi_1,v_{22}^*) +(u_{12},\varphi_2)\right ]
\Delta(\omega_2-\omega_{21}) + (u_{11},v_{21}^*)\Delta(2\omega_1-\omega_{21})
+(u_{12},v_{22}^*)\Delta(2\omega_2-\omega_{21}) +
$$
$$
+\left. \left [ (u_{11},v_{22}^*) +(u_{12},v_{21}^*)\right ]
\Delta(\omega_1+\omega_2-\omega_{21}) + \left [ (u_{12},u_{21}) +
(v_{22},v_{11})\right ] \Delta(\omega_2 - \omega_1 -\omega_{21})
\right\} \; ,
$$
$$
c_2^{(\ref{org2})} = c_2^{(0)}(t) + c_1^{(0)}\left\{ \left [
(\varphi_2,u_{11})+(v^*_{21},\varphi_1)\right ] \Delta(\omega_1-\omega_{21}) +
\right.
$$
$$
+ \left [ (\varphi_2,u_{12}) +(v^*_{22},\varphi_1)\right ]
\Delta(\omega_2-\omega_{21}) + (v^*_{21},u_{11})\Delta(2\omega_1-\omega_{21})
+(v^*_{22},u_{12})\Delta(2\omega_2-\omega_{21}) +
$$
\begin{equation}
\label{org71}
+\left. \left [ (v_{21}^*,u_{12}) +(v^*_{22},u_{11}) \right ]
\Delta(\omega_1+\omega_2-\omega_{21}) + \left [ (u_{21},u_{12})+
(v_{11},v_{22}) \right ]\Delta(\omega_2-\omega_1-\omega_{21}) \right\} \; .
\end{equation}

Expressions (\ref{org70}) show that one alternating field, with a frequency $\omega$,
can induce transitions between two topological modes, provided that one
of the following equations is valid:
\begin{equation}
\label{org72}
\omega=\omega_{21} \; , \qquad 2\omega=\omega_{21} \; .
\end{equation}
And Eqs. (\ref{org71}) demonstrate that two alternating fields, with frequencies
$\omega_1$ and $\omega_2$, can realize intermode transitions under one of the
resonance conditions:
$$
\omega_1=\omega_{21}\; , \qquad \omega_2=\omega_{21} \; ,
$$
$$
2\omega_1=\omega_{21} \; , \qquad 2\omega_2 =\omega_{21} \; ,
$$
\begin{equation}
\label{org73}
\omega_1+\omega_2 =\omega_{21} \; , \qquad
\omega_2 - \omega_1 = \omega_{21} \; .
\end{equation}
Going to the higher-order approximations in the iterative procedure
(\ref{org62}), we see that, in addition to the standard resonance conditions
as $\omega=\omega_{21}$ or $\omega_i=\omega_{21}$, there appear the conditions of
{\it harmonic generation}
\begin{equation}
\label{org74}
k\omega = \omega_{21} \qquad (k=2,3,\ldots)\; ,
\end{equation}
when there is the sole alternating field, or the conditions
of {\it parametric conversion}
\begin{equation}
\label{org75}
\sum_j (\pm \omega_j)  =\omega_{21} \; ,
\end{equation}
if several alternating fields modulate the trapping potential.

\begin{figure}
\centerline{
a) \includegraphics[width=7cm]{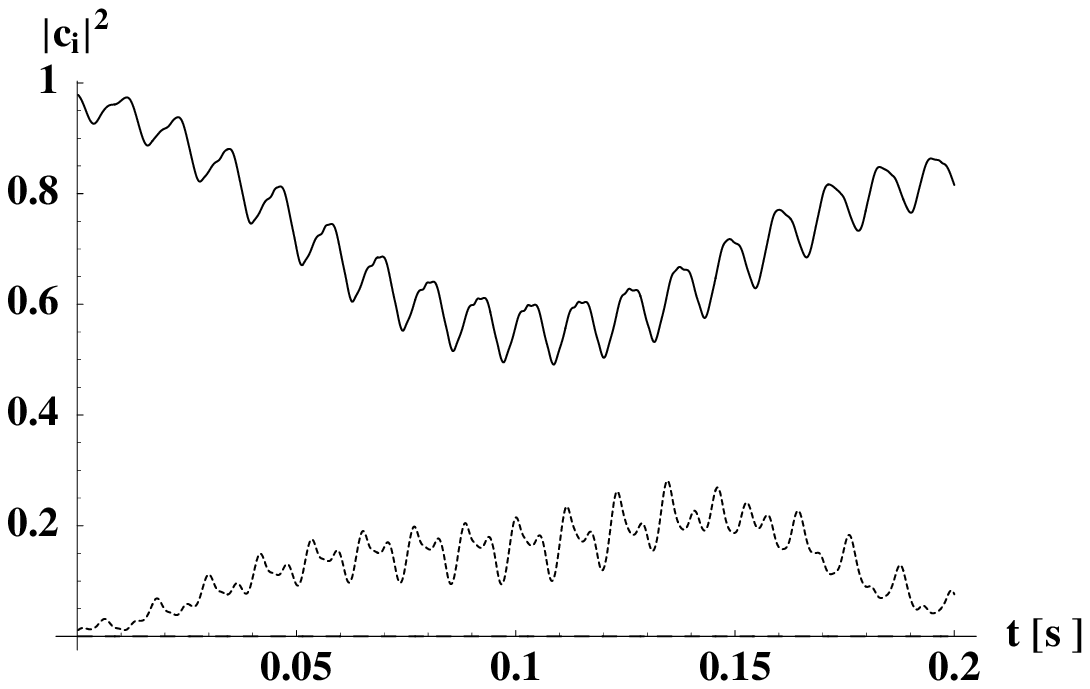}
b) \includegraphics[width=7cm]{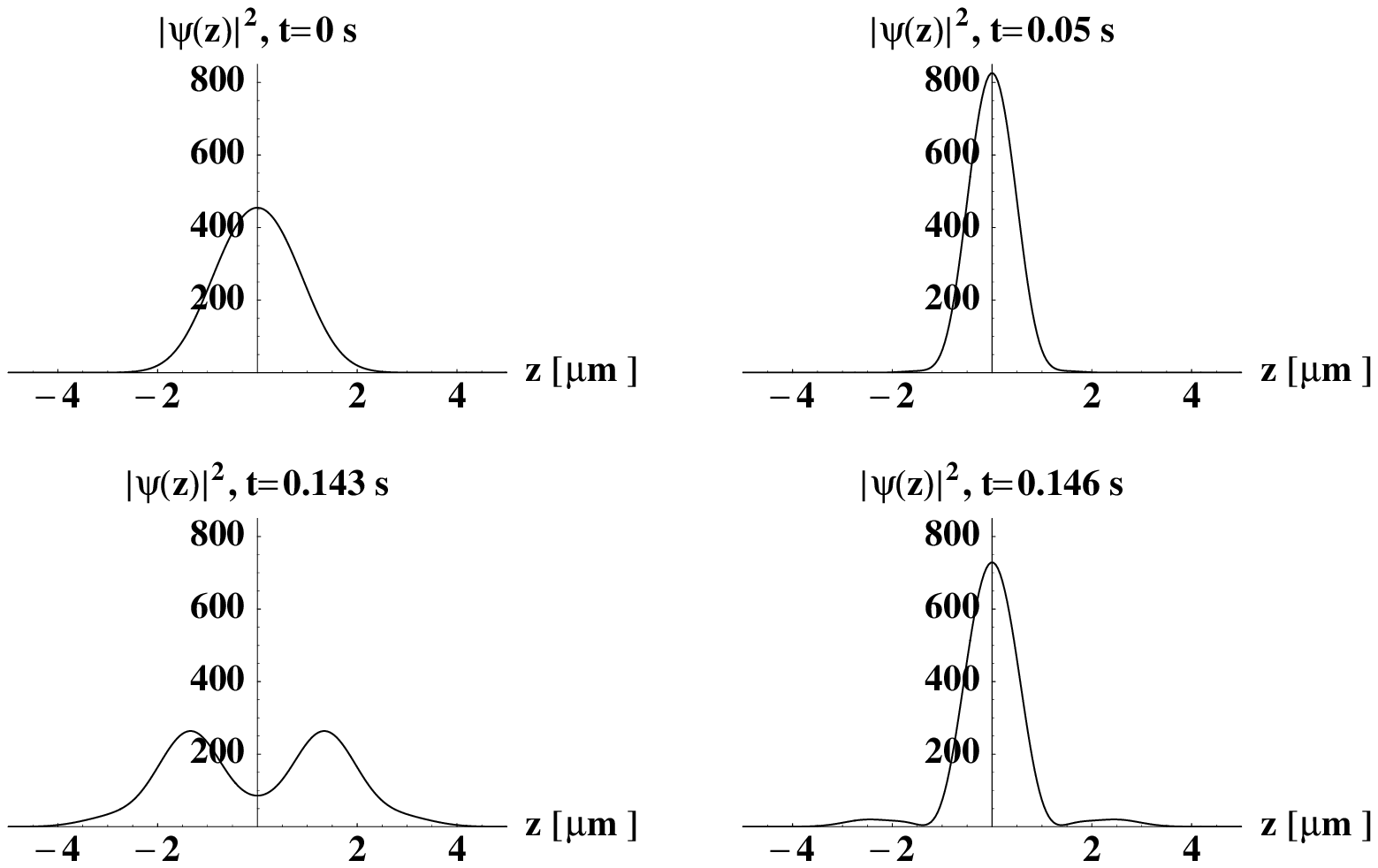}
}
\caption{\label{fig-harmgen} Time evolution of a BEC trapped and driven
by harmonic potentials.  a) mode coefficients averaged over a time of 5.3 ms
(the lines have the same meaning as in
Fig.~\ref{fig-anharmPotSubcrit}),
b) spatial probability density at different
time steps. The effect of harmonic generation of population in
the second excited state can clearly be seen in
both figures.}
\end{figure}

The prediction of nonlinear harmonic generation using the
averaging method is in excellent agreement with a respective
numerical simulation for a harmonic
potential of the form $U_z(z) = m_0 \omega_z^2 z^2 /2$, with
$\omega_z = 600$/s. To demonstrate harmonic generation
we employed a harmonic driving field of the form
(\ref{org6}) with $V_1(z) = \beta z^2$ and driving frequency
$\omega_1 = (E_2-E_1)/(2\hbar) \approx 552$/s. For a sufficiently
strong driving field ($\beta = 1.1\times 10^{-32}$ J/$\mu$m$^2$, roughly
half the strength of the trapping potential) we observed a strongly
enhanced population of the second excited mode (excitation
of the first excited state is forbidden by parity conservation).
The result for the time averaged coefficients is shown in
Fig.~\ref{fig-harmgen} and demonstrates a population of up
to 30 \% in this mode, far more than is to be expected in
absence of interaction. As in the case of an anharmonic
driving field (see above) we also find a substantial (30 \%) excitation of
other modes for a strong driving field.

The terms harmonic generation and parametric conversion are
used here by analogy with nonlinear optics, where there exist
analogous phenomena \cite{orgcit45}. Similar effects occur as well for
elementary excitations of Bose-Einstein condensates
\cite{orgcit48,orgcit49,orgcit50}.
As we have shown, such effects may also arise for topological
coherent modes.

In conclusion, the resonant generation of topological modes
seems to provide a feasible mechanism for creating novel
states of trapped Bose gases, containing the combinations
of several such modes. Here, we have considered the resonant
generation realized by modulating the trapping potential.
Note that another possibility of imposing resonant perturbations
could be done by inducing periodic variations in the atomic
scattering length \cite{adhikari03a,adhikari03b}.
The latter case requires that
the atoms are in a state close to a Feshbach resonance.

As we mentioned in the Introduction, a dipole topological mode was
successfully generated in experiment [23]. The resonant method of
vortex creation was analysed in detail in \cite{orgcit9}-\cite{orgcit11}.
Thus, the possibility of generating a single topological mode is well
justified. The mode can be of any nature, whether dipole or vortex.
The method of resonant generation works for any mode. Our numerical
simulations for the GPE, with realistic physical
parameters, show that the simultaneous generation of several topological
modes is also feasible.

\vskip 5mm

{\bf Acknowledgement}: This work was supported
by the Heisenberg-Landau Program, Alberta's informatics Circle of
Research Excellence (iCORE), 
the Forschergruppe Quantengase, and the Optik Zentrum Konstanz.
V.I.Y.  wishes to thank J\"urgen Audretsch for kind hospitality
during his stay in Konstanz.

\begin{appendix}
\section{Appendix. Description of numerical algorithms}
\label{sec:app}
Our numerical simulations are performed for a BEC that obeys
the GPE with an external potential of the form
$ U({\bf r}) = U_\perp({\bf r}_\perp) + U_z(z)$, i.e., a
transverse and a longitudinal trapping potential. 
We consider a quasi one-dimensional situation for which the
energy required to create excited states of the transverse trapping
potential $ U_\perp({\bf r}_\perp)$ is much higher than any other
energy. The wavefunction then takes the approximate form
$\varphi({\bf r},t) = \varphi_\perp({\bf r}_\perp)\; 
\psi(z,t)$, where $\varphi_\perp({\bf r}_\perp)$
is the transverse ground state. By integrating over the transverse
coordinates ${\bf r}_\perp$ one can show that $\psi$
fulfills a one-dimensional GPE of the form 
\begin{equation}
i\hbar\; \frac{\partial\psi}{\partial t} = \left ( 
\frac{\hbar^2}{2m_0}\;
\frac{\partial^2}{\partial z^2} + U_z(z) + N\; A_z\; |\psi|^2
+ \hat V\right )\; \psi \; ,
\label{1dgpe}
\end{equation}
with an effective coupling parameter
$A_z \equiv A_s \int |\varphi_\perp({\bf r}_\perp)|^4\;d{\bf r}_\perp$, 
whereby the integral over the transverse ground state is of the order 
of the transverse width squared.
The details of this reduction can be found in review
\cite{orgcit2}. 

In this appendix we shall present the methods used
for a direct numerical
solution of the one-dimensional GPE for different situations discussed
in the preceding sections.
We generally consider a BEC of 1000 $^{87}$Rb atoms
($m_0 = 1.45\times 10^{-25}$ kg, $a_s = 5.4$ nm) with a transverse
width of 7 $\mu$m. Our 1D simulation 
described a time evolution of 500 ms using $10^7$ discrete time
steps. It was performed on a grid
of 512 spatial points extending over a range of 26 $\mu$m.

\subsection{Numerical simulation of time evolution and determination
of nonlinear coherent modes}

To simulate the time evolution governed by Eq. (\ref{1dgpe}) we
have employed the well-known split-step Fourier method
\cite{deVries86}.
As initial condition we used the ground-state nonlinear coherent mode
$\varphi_0(z)$ associated with the potential $U_z(z)$ in absence of a
driving field $V(z,t)$. This ground state as well as higher nonlinear
coherent modes were numerically determined by an imaginary-time
propagation followed by a self-consistent field (SCF) method.

The imaginary-time propagation was described, for instance, in Ref.
\cite{dalfovo97}. Very roughly speaking it is a method to obtain
a kind of a small-temperature (corresponding to large imaginary time)
collective wavefunction which approaches the ground state wavefunction
for very large times. It usually gives excellent results for the ground
state but fails to reproduce excited nonlinear coherent modes unless
certain symmetry requirements are fulfilled. We used this method for
intermediate imaginary times to obtain a better trial wavefunction
$\varphi^{(0)}$ for the following SCF procedure. This step is not
necessary but helps to improve the speed of the numerical algorithm.

The SCF method consists in inserting the trial wavefunction
$\varphi^{(0)}$ into the nonlinear interaction part of Eq. (\ref{1dgpe}).
It thus provides a kind of potential term and turns Eq. (\ref{1dgpe})
into a linear equation for $\psi$. We then solved for the stationary
eigenstates of this equation. One of these states is then chosen to
become a new trial wavefunction $\varphi^{(1)}$ and then the procedure
is iterated until the overlap between $\varphi^{(n)}$ and
$\varphi^{(n+1)}$ deviates by less than $\varepsilon$ from 1 (we have
chosen $\varepsilon = 10^{-14}$). The choice of the appropriate new
trial wavefunction depends on the problem at hand. In the case of
a 1D Schr\"odinger equation with a confining potential, the eigenstates 
are real and the number $N_0$ of zeros of each nonlinear coherent mode  
turned out to be a good criterion for the next trial wavefunction. To 
determine the ground state we picked the  $N_0=0$ eigenstate in each 
iteration. For the excited coherent modes $N_0=1$ and 2 was the respective  
choice.

The convergence of the SCF method is the slower the higher the excited
state one is looking for, and it even can fail to converge for
$N_0 \geq 2$. We therefore implemented a convergence acceleration
scheme based on Anderson mixing as described in Ref.~\cite{eyter96}.
In short, this scheme enhances the usual SCF algorithm by adding
a kind of memory in the sense that $\varphi^{(n+1)}$ may depend
not only on $\varphi^{(n)}$ but on previous trial wavefunctions, too.
In practice, this memory lasts for 3-6 iterations. 
To implement this scheme one first writes $\varphi^{(n+1)}$ as a 
superposition of the previous trial wavefunctions.
Then one minimizes the deviation of this 
function from the previous trial wavefunction with respect to
the superposition coefficients. This leads to a simple
linear system of equations for the coefficients of the superposition.
Anderson mixing can speed up convergence by orders of magnitude
and turned out to be very suitable for our case. The three
nonlinear coherent modes of lowest energy for a harmonic potential and
an anharmonic potential are displayed in Fig.~\ref{fig-modes}.

We have checked if the nonlinear coherent modes obtained in this
way are indeed stationary solutions of the nonlinear
Schr\"odinger equation by propagating them for a sufficiently
long time using the split step algorithm. All modes were found
to be perfectly stationary: their density does not change and
their phase remains spatially homogeneous apart from the sharp
phase jump by $\pi$ that appears at the zeros of the excited
states, and apart from areas with extremely low densities
(less than about $10^{-13}$ of the peak density) where
numerical errors lead to a strongly fluctuating phase.

\subsection{Comparison to averaging procedure}

To compare the result of  the direct
numerical simulation with that obtained by the averaging procedure
we first projected the numerical solution $\psi(z,t)$ found using the
split-step method on the respective nonlinear coherent mode to obtain
as an intermediate result the non-averaged coefficients
$\tilde{c}_i(t_l) := (\varphi_i , \psi(t_l))\exp (i E_i t_l/\hbar)$ with
$ t_l = l \Delta t, l=0,1,2,\ldots$. Since
the nonlinear coherent modes $\varphi_i$ are not orthogonal
the sum $\sum_i |\tilde{c}_i(t_l)|^2$ is not unity and the individual
coefficients are oscillating on short time scales.

To obtain the coefficients $c_i(t_l)$ corresponding to the
averaging procedure we stored the wavefunction
at a total number of $N_t$ time steps
($N_t$ was typically on the order of 1000)  and defined the averaged
coefficients as
\begin{equation}
c_i(T_n) := \frac{1}{N_a} \sum_{l=1}^{N_a} \tilde{c}_i(T_n+t_l)\; ,
\end{equation}
where $N_a$ is the number of steps we averaged over and
$T_n := N_a n \Delta t$, $n=0,1,2,\ldots$. Generally $N_a=10$ to
$N_a=32$ resulted in a fairly smooth time evolution of the coefficients
$c_i(t)$ which can directly be compared to the results of the averaging
procedure.
\end{appendix}

\newpage

\end{document}